\documentclass[useAMS,usenatbib]{mn2e}
\usepackage{epsfig}
\def\lsim{\,\lower2truept\hbox{${<\atop\hbox{\raise4truept\hbox{$\sim$}}}$}\,}
\def\gsim{\,\lower2truept\hbox{${> \atop\hbox{\raise4truept\hbox{$\sim$}}}$}\,}

\title[The AT20G Bright Source Sample]{The Australia Telescope 20\,GHz (AT20G) Survey:\\The Bright Source Sample}
\author[Massardi et al.]{
\parbox[t]{\textwidth}
{Marcella Massardi$^{1,2}$\thanks{E-mail: massardi@sissa.it},
Ronald D.\ Ekers$^2$, Tara Murphy$^{3,4}$, Roberto Ricci$^5$,
Elaine M.\ Sadler$^3$, Sarah Burke$^6$, Gianfranco De
Zotti$^{7,1}$, Philip G.\ Edwards$^2$, Paul J.\ Hancock$^3$, Carole
A.\ Jackson$^2$, Michael J.\ Kesteven$^2$, Elizabeth Mahony$^3$,
Christopher J. Phillips$^2$, Lister Staveley--Smith$^8$, Ravi
Subrahmanyan$^9$, Mark A.\ Walker$^{10}$, and Warwick E.\
Wilson$^2$}
\vspace*{8pt} \\
$^{1}$SISSA/ISAS, Via Beirut 2--4, I-34014 Trieste, Italy\\
$^{2}$Australia Telescope National Facility, CSIRO, P.O.\ Box 76,
Epping, NSW 1710, Australia \\
$^{3}$School of Physics, University of Sydney, NSW 2006, Australia\\
$^{4}$School of Information Technologies, University of Sydney, NSW 2006, Australia\\
$^{5}$Department of Physics and Astronomy, University of Calgary,
2500 University Drive NW Calgary, AB, Canada\\
$^{6}$Swinburne University of Technology, P.O.\ Box 218, Hawthorn, Vic 3122, Australia\\
$^{7}$INAF, Osservatorio Astronomico di Padova, Vicolo
dell'Osservatorio 5, I-35122 Padova, Italy \\
$^{8}$School of Physics,University of Western Australia, 35 Stirling Highway Crawley, WA 6009, Australia\\
$^{9}$Raman Research Institute, Sadashivanagar, Bangalore 560080, India\\
$^{10}$Manly Astrophysics Workshop Pty Ltd, 3/22 Cliff St., Manly 2095, Australia}
\begin{document}

\date{}

\pagerange{\pageref{firstpage}--\pageref{lastpage}} \pubyear{2002}

\maketitle

\label{firstpage}

\begin{abstract}
The Australia Telescope 20\,GHz (AT20G) Survey is a blind survey
of the whole Southern sky at 20\,GHz (with follow-up observations
at 4.8 and 8.6\,GHz) carried out with the Australia Telescope
Compact Array (ATCA) from 2004 to 2007.

The Bright Source Sample (BSS) is a complete flux-limited
sub-sample of the AT20G Survey catalogue comprising 320
extragalactic ($|b|>1.5^\circ$) radio sources south of $\delta =
-15^\circ$ with $S_{20\rm GHz} > 0.50$\,Jy. Of these, 218 have
near simultaneous observations at 8 and 5\,GHz.

In this paper we present an analysis of radio spectral properties
in total intensity and polarisation, size, optical identifications
and redshift distribution of the BSS sources. The analysis of the
spectral behaviour shows spectral curvature in most sources with
spectral steepening that increases at higher frequencies (the
median spectral index $\alpha$, assuming $S\propto \nu^\alpha$,
decreases from $\alpha_{4.8}^{8.6}=0.11$ between 4.8 and 8.6~GHz
to $\alpha_{8.6}^{20}=-0.16$ between 8.6 and 20~GHz), even if the
sample is dominated by flat spectra sources (85~per cent of the
sample has $\alpha_{8.6}^{20}>-0.5$). The almost simultaneous
spectra in total intensity and polarisation allowed us a
comparison of the polarised and total intensity spectra: polarised
fraction slightly increases with frequency, but the shapes of the
spectra have little correlation. Optical identifications provided
an estimation of redshift for 186 sources with a median value of
1.20 and 0.13 respectively for QSO and galaxies.
\end{abstract}


\section{Introduction}

Our knowledge of the high frequency radio-source population is
poor. Important advances were made recently by the 15\,GHz
surveys with the Ryle telescope (Taylor et al.\ 2001; Waldram et
al.\ 2003) covering $520\,\hbox{deg}^2$ to a flux density limit of
25\,mJy and going down to 10\,mJy in small areas. The Wilkinson
Microwave Anisotropy Probe (WMAP) satellite has surveyed the whole
sky at 23, 33, 41, 61 and 94~GHz to completeness limits of $\gsim
1\,$Jy (Bennett et al.\ 2006; Hinshaw et al.\ 2007; L{\'o}pez-Caniego
et al.\ 2007), but there is little information in the flux
densities range between 200~mJy and 1~Jy.

High frequency surveys are very time-consuming. For telescopes
with diffraction limited fields of view the number of pointings
necessary to cover a given area scales as $\nu^2$. For a given
receiver noise, the time per pointing to reach the flux density
level $S$ scales as $S^{-2}$ so that, for a typical optically thin
synchrotron spectrum ($S\propto \nu^{-0.7}$), the survey time
scales as $\nu^{+3.4}$: a 20~GHz survey takes more than 110 times
longer than a 5~GHz survey with the same aperture covering the same
area of sky to the same flux density level.

However, Cosmic Microwave Background (CMB) studies, boosted by the
on-going NASA WMAP mission and by the forthcoming ESA Planck
mission, require an accurate characterization of the high
frequency properties of foreground radio-sources both in total
intensity and in polarisation. Radio sources are the dominant
contaminant of small-scale CMB anisotropies at mm wavelengths:
their Poisson contribution to temperature fluctuations is
inversely proportional to the angular scale, i.e. linearly
proportional to the  multipole number $l$, while the power spectra
of the CMB and of Galactic emissions decline at large~$l$. As a
result, Poisson fluctuations dominate for $l \gsim 400$. A high
frequency catalogue complete down to hundreds of mJy levels (the
rms in WMAP maps is $\gsim 200$~mJy at all frequencies) over a
large area, complemented with a good characterization of source
properties, can be used to correct the contaminating effect.

It may be necessary for the high frequency surveys to go even
fainter than the rms in the CMB observations. This is because
fluctuations in numbers of weak sources (particularly if there is
clustering) could affect power spectrum estimation: it depends on
the source counts at weak flux density levels and the contribution to the angular power spectrum
from sources at different flux density intervals.

Furthermore, forthcoming telescopes in the Southern hemisphere,
like the Atacama Large Millimeter Array (ALMA), that will operate
at frequencies above 90\,GHz, require suitable calibrators which
can be readily selected using large area high frequency surveys
(Sadler et al.\ 2006).

Optical surveys for transients, which would be severely
contaminated by Galactic novae, could use high frequency radio
surveys as templates to identify potential transients associated
with AGNs (Rau et al.\ 2007).

A Pilot Survey (Ricci et al.\ 2004; Sadler et al.\ 2006) at
18.5~GHz was carried out in 2002 and 2003 with the Australia
Telescope Compact Array
(ATCA\footnote{http://www.narrabri.atnf.csiro.au/}). It detected
173 objects in the declination range $-60^\circ$ to $-70^\circ$
down to 100~mJy.

The Pilot project characterised the high-frequency radio source
population and allowed us to optimise the observational techniques
for the full Australia Telescope 20~GHz (AT20G) Survey. The full
survey will cover the whole Southern sky to a flux density limit
of $\simeq 50$~mJy. It began in 2004 and will be completed in
2007. To date we have completed the survey from the South Pole to
declination $\delta = -15^\circ$. More than 4400 sources were
detected, down to a flux density of 50~mJy. We expect another 1500
sources in the declination range $-15^\circ<\delta<0^\circ$ for
which the analysis is still on-going.

This paper presents the analysis of the brightest ($S_{20\rm GHz}>
0.50$~Jy) extragalactic ($|b|>1.5^\circ$) sources  in the AT20G
Survey based on the observations in the declination range
$-90^\circ<\delta<-15^\circ$ surveyed between 2004 and 2007.

In \S~\ref{sec:Observations} we describe the observing techniques.
In \S~\ref{sec:Pipeline} we describe the data reduction methods
that have been applied to the whole survey and the selection
criteria for the Bright Source Sample. After presenting the sample
in \S~\ref{sec:Sample}, in \S~\ref{sec:Properties} we illustrate
its main properties in terms of source populations, spectral
behaviour in total intensity and polarised emission. We also
compare our results with those of analyses of samples at lower or
similar frequencies and discuss the possible impact on CMB
studies. In \S~\ref{sec:Conclusions} we summarise our conclusions.
In Appendix~\ref{sec:IndSrc} we list some notes on individual
sources of interest.

\section[]{Observations}\label{sec:Observations}
\begin{table*}
 \centering
  \caption{Follow-up observations to confirm candidate sources at 20~GHz (flagged as \emph{C}),
  to observe them at 5 and 8~GHz (\emph{O}) or to repeat previous bad quality observations
  (\emph{R}). (\emph{M}) refers to the observation in which we observed the very extended sources in
  mosaic mode.
   \label{tab:obstable}}
  \begin{minipage}{165mm}
  \begin{tabular}{@{}clccccc@{}}
\hline
\textbf{Epoch}& \textbf{Declination}& \tiny{\textbf{Central }} & \tiny{\textbf{Array Configuration }}        &\textbf{Beamsize}&  & \\
\textbf{ref.}      &\textbf{range}      & \tiny{\textbf{Frequencies(MHz)}} & \tiny{\textbf{(shortest spacing [m])}}&[arcsec]&  \textbf{Dates} & \textbf{Reason} \\
 \hline
1&$-50^\circ,-30^\circ$  &  18752, 21056  &   H214 (80)  &$10.7\times10.7$                & 21 Oct - 27 Oct 2004 & C   \\
1&$-50^\circ,-30^\circ$  &  4800, 8640    &   1.5C (77)  &$8.3\times12.8$ $4.6\times7.13$ & 04 Nov - 08 Nov 2004 & O   \\
2&$-90^\circ,-50^\circ$  &  18752, 21056  &   H168 (61)  &$13.9\times13.9$                & 27 Oct - 31 Oct 2005 & C   \\
2&$-90^\circ,-50^\circ$  &  4800, 8640    &   1.5C (77)  &$8.3\times8.8$ $4.6\times4.9$   & 12 Nov - 15 Nov 2005 & O   \\
3&$-90^\circ,-30^\circ$  &  18752, 21056  &   H214 (80)  &                                & 29 Apr - 03 May 2006 & R   \\
3&$-90^\circ,-30^\circ$  &  4800, 8640    &   1.5D (107) &                                & 19 Jun - 23 Jun 2006 & R,O \\
4&$-30^\circ,-15^\circ$  &  18752, 21056  &   H214 (80)  &$2.0\times5.1$                  & 14 Oct - 17 Oct 2006 & C   \\
4&$-30^\circ,-15^\circ$  &  4800, 8640    &   1.5B (30)  &$8.3\times21.1$ $4.6\times11.7$ & 09 Nov - 12 Nov 2006 & O   \\
5&$-90^\circ,-15^\circ$  &  18752, 21056  &   H214 (80)  &                                & 11 May - 16 May 2007 & R   \\
5&$-90^\circ,-15^\circ$  &  4800, 8640    &   1.5C (80)  &                                & 04 May - 10 May 2007 & R,O \\
6&$-90^\circ,-30^\circ$  &  16704, 19392  &   H75 (31)   &$35.3\times35.3$                & 01 Oct 2006 & M \\
\hline
\end{tabular}
\end{minipage}
\end{table*}

\subsection{Survey mode}

The first phase of our observations is to make a set of blind
scans. More details on the survey mode observations, the
mapmaking, the source detection in maps and the completeness
and reliability of the survey will be provided in forthcoming
papers on the full AT20G sample. Here we present only a general
description.

We have exploited the ATCA fast scanning capabilities (15~degrees
min$^{-1}$ in declination at the meridian) and the 8~GHz bandwidth
of the wideband analogue correlator originally developed as part of
the collaboration for the Taiwanese CMB experiment AMiBA (Lo et al.\ 2001)
and now applied to 3 of the six 22~m dishes of the ATCA. The
lag-correlator measures 16~visibilities as a function of
differential delay for each of the three antenna pairs used. This
wideband analogue correlator has no mechanism to allow for
geometrical delay as a function of the position in the sky, so the
scan has to be performed along the meridian corresponding to zero
delay for the EW configuration used. There is no fringe stopping.

The scanning strategy consists of sweeping sky regions 10$^\circ$
or 15$^\circ$ wide in declination, using a whole Earth rotation to
cover all the right ascensions in a zig-zag pattern. Each
declination strip requires several days to be completely covered
by moving the scanning path half a beam apart from day to day.
Along the scan a sample is collected every 54~ms (3~samples per
beam), enough to reach a rms noise of 12~mJy. With this
exceptional continuum sensitivity, along with precise and high
speed telescope scanning capability, we can scan large areas of
the sky, despite the small ($\sim 2.4$~arcmin) field of view at
20~GHz. Scans with bad weather or occasional equipment error have
been repeated, so that the sky coverage is 100 per cent at the
flux density levels we are interested in this paper.

Candidate sources are identified by looking for telescope response
pattern within the delay channels in the time ordered data,
correlated between the baselines. The correlator outputs for each
set of 24-hour observations were interleaved and calibrated to
produce maps. The overall rms noise in the maps reaches $\simeq
10$~mJy.

The initial calibration is based on a transit observation of a
known calibrator observed every 24~hours between scans. All the
sources detected in the scans that have known flux densities and
positions (about 10 for each scan) are then used in a bootstrap
process to refine the scan calibration. From this we produced an
initial list of positions and flux densities for candidate sources
brighter than $5\sigma$\ (about 50~mJy).

\subsection{Follow-up mode}

Each of the candidate sources selected in the first phase has been
re-observed to confirm they are genuine sources and to get
accurate positions, flux densities and polarisation information.

Note that this procedure will exclude any fast (within few weeks)
transient sources, if they exist. We intend to check for such
objects in a future analysis. The follow-up has been performed
with an hybrid array configuration (i.e., with some of the baselines on
the NS direction) with the normal ATCA digital correlator with two
128~MHz bands centered at 18752~MHz and 21056~MHz and two
polarisations. The combination of the two close bands could be
considered as a single 256~MHz wide band centered at 19904~MHz,
which is the reference frequency for our `20~GHz' observations.

The follow-up observations exploit the fast mosaic capabilities of
the ATCA to reduce the slewing time between pointings. In our
observing strategy each mosaic point is a pointing on a candidate
source. The same source has to be observed more than once to
improve the visibility plane coverage. The sources discussed in
this paper have been observed at least twice and in some cases up
to 8 times at different hour angles. Up to 500~candidates could be
followed-up in a day. A set of secondary calibrator sources are
regularly observed between blocks of candidate sources.

Within a couple of weeks, we observed the confirmed sources with
an East-West extended array configuration with two 128~MHz bands
centered at 4800~MHz and 8640~MHz to study their radio spectral
properties. Those are the frequencies to which we will refer in
the following as `5' and `8'~GHz. In Table~\ref{tab:obstable} we
have summarised the array configurations used to observe sources
or to replace previous bad quality data in the various sky
regions. The simultaneity of observations at different frequencies
is necessary to study the spectral properties of the sources,
avoiding errors due to the source variability.

The primary beam FWHM is $2.4$,  $5.5$, and $9.9$~arcmin at 20, 8,
and 5~GHz, respectively.

We carried out observations dedicated to high sensitivity
polarisation measurements in October~2006 with the ATCA on a
sub-sample of the bright sources, using the most compact
configuration, H75 (Burke et al., in preparation). This provided
more accurate short-spacing measurements of flux densities at
20~GHz, imaging and integrated flux densities. Nine very extended
sources have been selected from low frequency catalogues (PMN,
Griffith \& Wright\ 1993, and SUMSS, Mauch et al.\ 2003, 2007) to
be observed in mosaic mode to improve the flux density estimation
at 20~GHz. Some of those values will be used here (as explained
below) in order to avoid flux density underestimation due to
resolution effects. The lack of low frequency data with the same
imaging mosaic observations to get integrated flux densities does
not allow us to use these data for spectral analysis, even though
they have a more precise determination of integrated flux
densities at 20~GHz.

\section[]{Data reduction}\label{sec:Pipeline}

\begin{figure}
\begin{center}
\includegraphics[width=8cm]{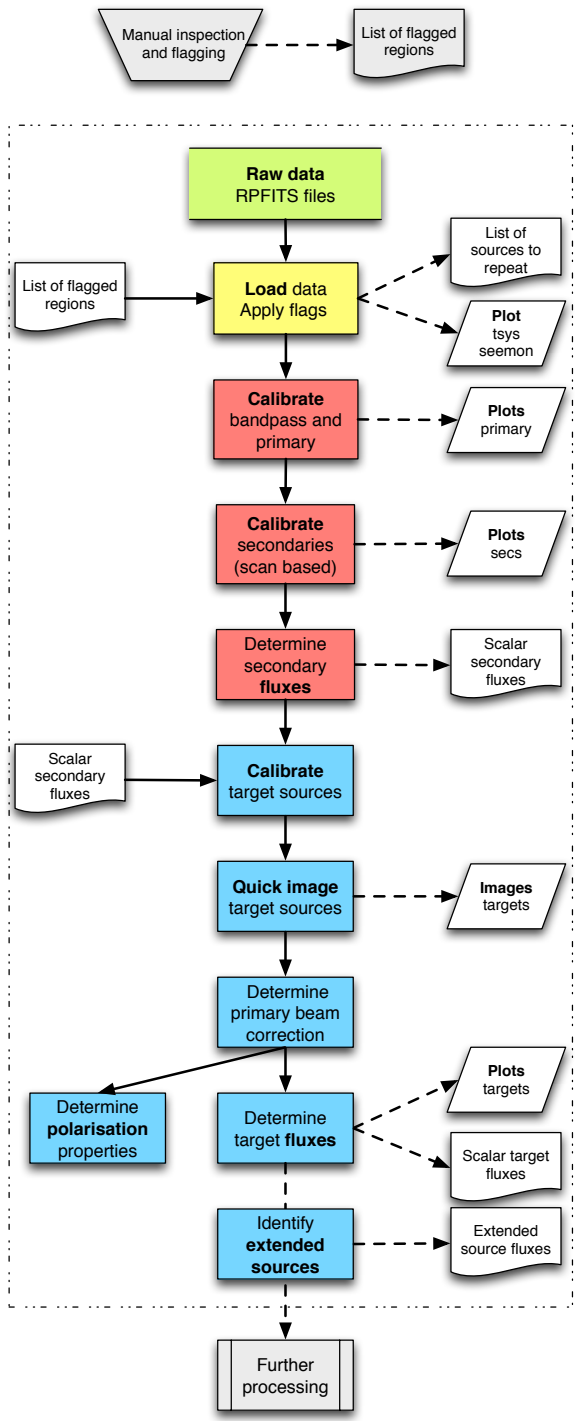}
\caption{Diagram of the analysis pipeline
process.}\label{fig:calibration}
\end{center}
\end{figure}

We have developed a fully automated custom analysis pipeline to
edit, calibrate, and reduce the data for all the follow-up
observations (Fig.~\ref{fig:calibration}). This procedure has been developed to ensure
consistent data quality in the final catalogued data. The software
was built using the scripting language Python, and the underlying
data reduction was done with the aperture synthesis reduction
package Miriad (Sault, Teuben \& Wright 1995).

After an initial manual inspection of the data to flag bad data,
the pipeline generates the calibration solutions. Once source flux
densities are calibrated, a set of processes is applied to
determine positions, peak flux densities, extendedness, integrated
flux densities, polarisation properties and to generate images.
The final result is a list of confirmed sources with all the
available information and images for each epoch and for each
frequency. From this list we have selected the Bright Source
Sample that we will analyze in the following sections.

In the rest of this section we describe the details of the data reduction.

\subsection{Flagging of bad data}

An initial inspection of the correlator output is necessary in
order to identify interference or any problems in the data acquisition
that mean the data should be excluded from further analysis.

Weather conditions can seriously affect the quality of the data.
Attenuation of the signal by atmospheric water vapour can decrease
the sensitivity of the observations, and atmospheric turbulence
can produce phase fluctuations that may produce visibility
amplitude decorrelation. Data collected in periods of bad weather
have to be removed. In particular, calibrator data must be of high
quality otherwise it introduces errors in the calibration
solutions that affect the whole dataset (Thompson, Moran \& Swenson 2001).

A seeing monitoring system is run at the ATCA site simultaneously
with the main array. Two 40~cm dishes on a 240~m baseline monitor
the differential phase variations in a geostationary satellite
signal caused by tropospheric water vapour fluctuations. These
fluctuations can be used to estimate the decorrelation in the
interferometric data (Middelberg, Sault \& Kesteven 2006). In
addition, the absorption due to atmospheric water vapour is
estimated for each main antenna receiver by measuring the system
temperature ($T_{\rm sys}$) changes due to tropospheric emission.

We used the seeing monitor data (to measure amplitude
decorrelation) in conjunction with $T_{\rm sys}$ (to estimate
tropospheric opacity) to develop semi-automatic flagging criteria.
Specifically, we discarded data from all the periods in which
there was decorrelation greater than 10~per cent. This improves
the uniformity and data quality across all our observing epochs.

Flux density measurements for unresolved target sources suffering from
significant decorrelation could still be recovered using triple product
techniques (see below), but imaging for these sources was not possible.
Calibrators with significant decorrelation were excluded, and the blocks
of target sources associated with those calibrator observations were also
excluded.
Very occasionally, bad weather required large blocks of data to be edited
out and hence a small number of sources do not have near-simultaneous data
at the lower frequencies (5 and 8~GHz).

\subsection{Calibration}\label{sec:Calibration}

Primary flux calibration and bandpass calibration were carried out
in the standard way using PKS~B1934$-$63 as the primary and
PKS~B1921$-$293 as the bandpass calibrator.

For the secondary flux calibration we follow a non-standard procedure,
which we describe here.
Our follow-up observing schedule follows the pattern:
\begin{itemize}
\item a nearby secondary calibrator is observed for $\sim 5$
minutes \item a block of  $\sim 20$ target sources are observed
for $\sim 40$ seconds each \item the secondary calibrator is
re-observed for $\sim 5$ minutes
\end{itemize}
This pattern is repeated throughout the observations. Hence we
typically observed around $\sim 50$ secondary calibrators during
one epoch of our observations. To calculate an accurate flux
density for each secondary calibrator we calculate the mean of the
individual snapshot flux densities across the whole run excluding
only a snapshot which has a flux density more than 2 standard
deviations away from the mean. The rest of the snapshots are
averaged to calculate the flux density for that secondary.
Finally, each target source is calibrated using the secondary
calibrator associated with its observing block. For each target
source we calculate the position, flux density, primary beam
corrections and Stokes parameters.

Full polarisation data ($I$, $Q$, $U$ and $V$ Stokes parameters)
are determined for all of the target sources. These are calculated
in the pipeline using a polarisation specific process. Firstly, a
correction is applied for the time dependent phase difference
(automatically monitored in real time at the telescope) between
the orthogonal, linear antenna feeds (which are referred to as x
and y). After making this correction, a small residual xy-phase
signal still remains. Because we have insufficient secondary
calibrator data to accurately determine all the free parameters
involved in instrumental polarisation corrections (e.g., leakages,
residual xy-phase difference, and time-dependant gains), leakage
terms were calculated using the primary calibrator,
PKS~B1934$-$638. The linear polarisation of this calibrator is
known to be not variable and less than 0.2~per cent of the total
source flux density at each of our observing frequencies. To
determine the leakage terms, it was assumed to be unpolarised. We
copied the leakage values to all the secondary calibrators,
simultaneously calculating the time-dependent complex antenna
gains, the residual xy phase differences, and the $Q$ and $U$
Stokes parameters of the calibrators. The polarisation calibration
was then applied to the target sources.

\subsection{Extended Sources}

If a source is extended more than few arcsec (depending on the
array configuration) we will underestimate its total flux density
using either the image peak or the triple correlation. We could
use the shortest spacing or integrate the image over a larger area
to recover the total flux density for an extended source, but this
does not optimise the sensitivity for a point source. Hence we
need an automatic procedure capable of distinguishing point-like
sources from extended sources. To do this we exploited the
properties of the observed phase closure. The phase closure
calculated on three antennas (a baseline closure triangle) is the
vector combination of the phase of the correlated signal between
each couple of antennas:
\begin{equation}\label{eq:phcl}
\Phi_{\rm cl}= \Phi_{1,2}+\Phi_{2,3}-\Phi_{1,3}.
\end{equation}
It is null for a point source. It is also null for any flux
density distribution that is an autocorrelation function such as a
symmetrical Gaussian, but this is unlikely to occur for the sample
of extragalactic objects we are considering in this paper.

In an array with more than three antennas the root mean square
(rms) of the phase closure can be calculated for all the possible
combinations of three antennas in the array. Analogously to the
three antenna case, it is expected to be null for a point source:
the phase closure rms is different from 0 if the source is
extended or if there is more than one source in the beam area.
Receiver noise will contribute to the phase closure errors but the
phase closure rms does not depend on antenna based instrumental
and atmospheric phase effects or on the position of the source in
the field.

For each source we compare the observed phase closure to the
predicted phase closure due to receiver noise.  This is determined
by Monte Carlo simulations of our observations for point sources
with receiver noise added.

Then we have defined the \emph{extendedness parameter} as the
ratio of the predicted phase closure rms due to noise and the
observed value. The discrimination between point-like and extended
sources is for the extendedness parameter equal to 3, a good trade
off, minimizing the wrong assignments to the two classes. An
incorrect assignment will result in a flux density error of at
most 20 per cent passing from one class to another. The largest
errors are made for faint objects (well below $0.50\rm\,mJy$).

With the 214\,m array the threshold means that a source is
extended if it has significant flux density ($>10$ per cent) at
20~GHz on scales larger than 6 arcsec.

The same criterion could be applied to all the frequencies, but,
in the following, we consider that a source is extended if its
extendedness parameter is larger than 3 at 20~GHz. A more refined
method will be required to correct for confusion due to faint
sources especially at 5~GHz, but this correction is negligible for
the present sample.

\subsection{Source Positions}\label{sec:position}

Source positions have been measured on the source centroid of the
cleaned and restored images. Formal positional errors in right
ascension and declination have been obtained by quadratically
adding a calibration term ($\sigma_{\rm cal}$) and a noise term
($\sigma_{\rm n}$). We have statistically determined the
calibration term by cross-matching the Bright Source Sample (233
observations in different epochs) with the International
Coordinate Reference Frame catalogue (ICRF, Ma et al.\ 1998). The
VLBI-measured positions in the ICRF catalogue are accurate to $\le
10^{-3}\,$ arcsec, so any discrepancy between the positions of our
target sources and the ICRF positions can be attributed to
positional errors in our sample. The rms positional error is
0.5~arcsec in right ascension and declination with small
variations due to changing weather conditions. For the Bright
Source Sample the noise term is always negligible.

\subsection{Flux Density Measurements}

We have obtained the flux densities for bright point-like sources
using the triple product method implemented in the Miriad task
CALRED. The amplitude of triple product is the geometric average
of the visibility amplitudes in a baseline closure triangle
\begin{equation}
A_{\rm TP}=\sqrt[3]{A_{1,2}\cdot A_{2,3}\cdot A_{3,1}}
\end{equation}
and its phase is the phase closure (eq.~\ref{eq:phcl}).

This way of measuring flux densities is particularly well suited
for strong and point-like sources and it is able to recover the
flux density lost in imaging because of phase decorrelation. We
have derived formal flux density errors, by adding quadratically a
calibration term (gain error, $\sigma_{\rm gain}$) and a noise
term ($\sigma_{\rm n}$). The gain error is a multiplicative term
(i.e., it is proportional to the source flux density) and is a
measure of the gain stability over time. We estimated $\sigma_{\rm
gain}$ for each observational epoch and frequency from the scatter in
the visibility amplitudes of the calibrators in each
observing run. Such average values for the gain errors were found
to be of the order of a few per cent. The noise term is an
additive term strictly related to the interferometer noise which
is proportional to the system temperature. Since no source has
significant Stokes $V$, the rms noise levels in the $V$ images
have no gain error and are used as an estimate of the $\sigma_{\rm
n}$ value for each target source.

For sources that have been defined as extended at 20~GHz,
integrated flux densities at 5, 8 and 20~GHz have been estimated
from the amplitude of the signal measured by the shortest
baseline. Any source extended at 20~GHz is assumed to be extended
at 5 and 8~GHz. Sources which are extended  at 5 or 8~GHz but
core-dominated at 20~GHz won't be considered as extended according
to this procedure. In this case we are assuming a dominant point
source and the flux densities at all the frequencies will be for
the core and not the total source. The shortest baseline used in
the follow-up (see Table~\ref{tab:obstable}) is 60 or 80~m so we
still underestimate flux densities for sources larger than
20~arcsec. For extended sources the error is increased by the
square root of the number of baselines $n_{\rm base}$ (normally 10 for our
5-antenna follow-up arrays) to correct for the fact that the flux
densities for these sources are estimated using only one (the
shortest) baseline instead of $n_{\rm base}$.

\subsection{Polarisation}\label{sec:polarisation}

Images in Stokes $U$, $Q$ and $V$ are calculated for all the
target sources using the calibration procedure described in
\S~\ref{sec:Calibration}. Since no sources have detectable $V$ at
our sensitivity the $V$ image is used to estimate the noise error.
If a source is detected, $P$, the polarised flux, is calculated in
the usual way $P = \sqrt{Q^2 + U^2}$  with no noise debias factor.
For the intensity ($I$) we were able to avoid the effect of phase
decorrelation by using the triple product but we don't have an
equivalent measure for $U$ and $Q$. However, the tropospheric
phase decorrelation affects Stokes parameters $Q$ and $U$ in
exactly the same way as Stokes $I$, so that we can use the triple
product amplitude, $I_{\rm tp}$, and the restored Stokes $I$ image
peak, $I_{\rm map}$, to calculate the factor by which the flux
density is reduced due to decorrelation $\chi = I_{\rm
map}/I_{tp}$. Then the corrected polarised flux is $P/\chi$.

The error on $P$ is $P_{\rm ERR}= \sqrt2\sigma_V/\chi$, where
$\sigma_V$ is the noise error from the $V$ image: that is the
propagation of the error on $P$ assuming that both the errors on
$U$ and $Q$ are equal to $\sigma_V$.

For the non-detections ($P<3\sigma_V$) we calculate an upper limit
on $P$ setting $U$ and $Q$ to $3\sigma_V$ and calculating the value of $P$
as above.

To avoid bias in $P$, it is always measured at the position of the
peak in $I$ for point sources. For extended sources we need to
integrate the polarisation vectors over the source which is the
same as the integrated value of $U$ and $Q$. This has been done
for the extended sources that have been observed in the mosaic
mode, but at this time we have not determined the integrated polarisation
for the slightly extended sources.

Unfortunately, an instrumental phase problem has spoiled the phase
measurements for May 2007 observations: for this epoch flux
densities could be recovered with triple correlation techniques,
but the polarisation information had to be flagged.

\section[]{The sample}\label{sec:Sample}

\begin{figure}
\begin{center}
\includegraphics[width=7cm, angle=90]{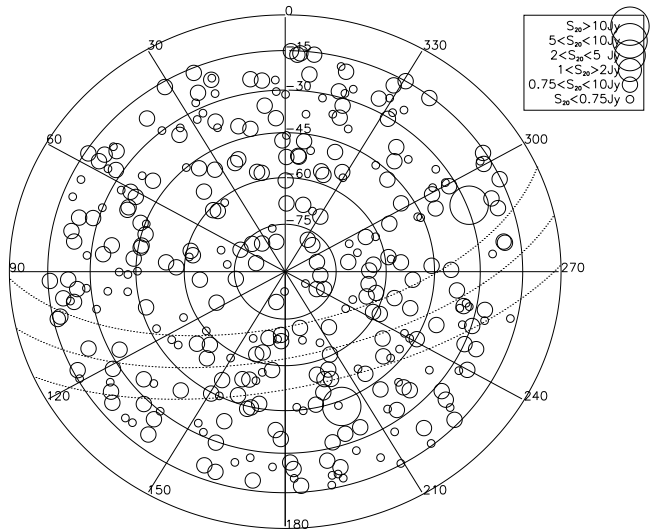}
\caption{Equal area projection of the Southern sky in equatorial
coordinates, showing the BSS sources. The symbols size is a
function of the flux density at 20\,GHz, as in the inset. The
dotted lines indicate the regions of Galactic latitude
$b=\pm10^\circ$ and the Galactic plane.}\label{fig:map}
\end{center}
\end{figure}

From the confirmed sources observed in the period 2004-2007 we
selected those which have good quality data at 20~GHz and flux
densities above 0.50~Jy and Galactic latitude $|b|>1.5^\circ$.
Some sources were observed at more than one epoch, in which case
the flux density selection threshold has been applied to the
measurements with the highest quality and the smallest primary
beam correction. To avoid any selection bias caused by
variability, sources were only included if they were above the
threshold for the epochs with the highest quality observation.
This is necessary to avoid any bias caused by variability. The
final distribution in coordinates, both equatorial and Galactic,
is homogeneous (Fig.~\ref{fig:map}). The median errors in flux
density estimation is 4.8~per cent at 20~GHz, and 2.5 and 1.5~per
cent respectively at 8 and 5~GHz.

A small number of very extended sources are known to have 20~GHz
flux density above our 0.50~Jy cut. These are discussed in
\S~\ref{sec:extended}.

If a simultaneous observation at low frequency is available and it
satisfies all our quality requirements we use it for the spectral
analysis. If no simultaneous data is available we report the best
low frequency observation we have, but that source will not appear
in our spectral analysis.

\subsection{Catalogue}

Tables\,2 and\,3 catalogue the 320 sources in the sample. Table\,2
lists positions, flux densities, identifications with other
optical or radio catalogues, and redshifts. Table\,3 lists the
information about polarisation (polarised flux densities,
fractions and angle of polarisation). For the full sample the
source names reflect the source J2000 equatorial coordinates as
`AT20G JHHMMSS-DDMMSS'. For sake of simplicity in this paper we
will refer to the sources according to their sequential number as
listed in the first column of Table~2.

The content of the columns are as follows for Table~2.
\begin{itemize}
    \item (1) Sequential number. An asterisk (`*') following
the number indicates that the source is listed in the
Appendix~\ref{sec:IndSrc} or has been commented on in the text.
    \item (2--3) Right ascension and declination (J2000). The
average error in right ascension and declination is 0.5 arcsec
(see \S~\ref{sec:position}).
    \item (4--5) Flux density at 20~GHz and its error in Jy.
    \item (6--7) Flux density at 8~GHz and its error in Jy.
    \item (8--9) Flux density at 5~GHz and its error in Jy.
    Whenever available we give the results of 5 and 8~GHz observations almost simultaneous to the
    20~GHz ones, otherwise we refer to the best observations available for the source at each frequency.
    \item (10--11) Flux density at 1.4\,GHz and its error from NVSS (Condon et al.\ 1998).
    \item (12--13) Flux density at 0.843\,GHz and its error from SUMSS (version 2.0).
    \item (14--15) Redshift and its reference, obtained as discussed in \S~\ref{sec:optid}.
    \item (16) Optical B magnitude for sources with SuperCOSMOS\footnote{http://www-wfau.roe.ac.uk/sss/} counterparts.
    \item (17) SuperCOSMOS identifications: `G' for galaxies, `Q' for
QSOs. A blank space indicates that no identification was possible
(see \S~\ref{sec:optid}).
    \item (18) Flags column where we collected some flags for source
properties in the following order:\begin{itemize}
\item the epoch
of the 20\,GHz observations: numbers refer to the epoch reference
number in Table \ref{tab:obstable};
\item spectral shape: `F' for
flat, `I' for inverted,`P' for peaked, `S' for steep,`U' for
upturning, as in Table \ref{tab:alphatable};
\item galactic position: a `G' indicates that
the source is within 10$^\circ$ from the galactic plane;
\item epoch of
observation at 8 and 5\,GHz respectively, in case of not
simultaneous observations (numbers refer to the epoch reference
number in Table \ref{tab:obstable}): in such cases we have listed the flux
densities measured in the best observation available.
\item
extendedness: `E' if the source is extended at 20~GHz, `M' if it
has been observed in the mosaic mode. The flux density for the `M'
sources corresponds to the integrated flux density of the source
in the mosaic area;  \item a
flag `C' means that the source is listed in the AT calibrator
manual \end{itemize}
   \item (19) Alternative name from other well known catalogues (PMN,
PKS) at radio frequency.
    \item (20) Identification number in the WMAP 1-yr catalogue (Bennett et al.\ 2003).
\end{itemize}

In Table~3 we collected the following columns
\begin{itemize}
    \item (1) Sequential number as in Table~2.
    \item (2--3) Right ascension and declination (J2000).
    \item (4--5) Integrated polarised flux in Jy and its error at 20~GHz.
    \item (6) Fractional polarisation at 20~GHz (per cent).
    \item (7) Polarisation angle at 20~GHz in degrees.
    \item (8--9) Integrated polarised flux in Jy and its error at 8~GHz.
    \item (10) Fractional polarisation at 8~GHz (per cent).
    \item (11) Polarisation angle at 8~GHz in degrees.
    \item (12--13) Integrated polarised flux in Jy and its error at 5~GHz.
    \item (14) Fractional polarisation at 5~GHz (per cent).
    \item (15) Polarisation angle at 5~GHz in degrees.
\end{itemize}

\subsection{Source counts} \label{sec:completeness}
\begin{figure}
\begin{center}
\includegraphics[width=6cm, angle=90]{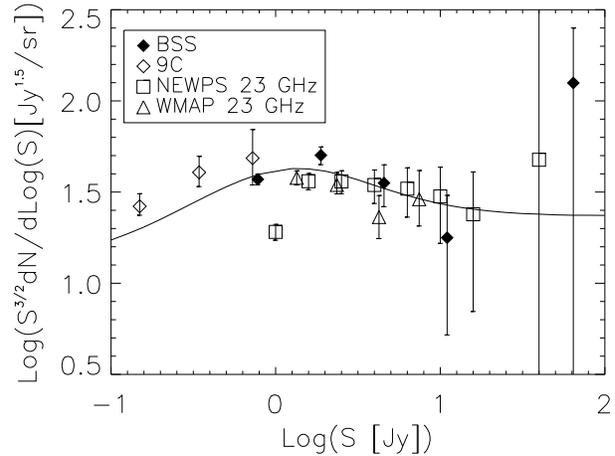}
\caption{Differential source counts at 20\,GHz, with their Poisson
errors, normalised to Euclidean counts. The statistics is very
poor above $\simeq 1\,$Jy. The model by De Zotti et al.\ (2005) is
also shown for comparison. Points from the 9C Survey (Waldram et
al. 2003), and from the catalogues based on WMAP maps have been
overlapped (WMAP, Hinshaw et al. 2007; NEWPS, L{\'o}pez-Caniego et
al.\ 2007).}\label{fig:20GHzsrccnt}
\end{center}
\end{figure}

The differential source counts for the present sample
(Fig.~\ref{fig:20GHzsrccnt}) are in good agreement with the 9C
counts at 15\,GHz (Waldram et al.\ 2003) and with the WMAP counts
at 23\,GHz (Hinshaw et al.\ 2007; L{\'o}pez-Caniego et al.\ 2007),
as well as with the predictions of the model by De Zotti et al.\
(2005). However, we must beware of resolution effects. The source
detection technique is optimised for point-sources, and there will
be some bias against extended sources with angular sizes larger
than about 30~arcsec. An outstanding case is Fornax~A, one of the
brightest sources in the Southern sky, which was missed by our
survey because its compact nucleus (and any other compact
component) is fainter than our blind scan detection limit (as was
expected according to previous observations, e.g., Morganti et
al.\ 1997) and its lobes are completely resolved by the 30-m
baseline used for the blind scan.

By the same token, although no other bright source appears to have
been completely missed by the AT20G Survey, the flux densities of
the most extended objects may fall below our threshold because
they are underestimated. To overcome this problem we have searched
low-frequency catalogues for bright and extended sources, expected
to have integrated 20~GHz flux densities above our 0.50~Jy
threshold but missed by our selection (see \S~\ref{sec:extended}).
For these sources we have made use of the information collected in
mosaic mode during the October~2006 polarisation follow-up run
(Burke et al.\ in preparation). Except for Fornax~A, all the known
bright extended sources in our area have been mosaiced.

Another source of uncertainty in the sample selection is
variability, making sources move in or out of a given flux density
bin, depending on the epoch of observations. Since we have been
gathering flux density measurements made at different times we do
not have a uniform view of the surveyed sky region. Only 30 BSS
sources have more than one observation at 20~GHz in the 2002--2007
period (considering also the Pilot Survey observations), too small
a sample for a meaningful analysis of variability. However, Sadler
et al. (2006) found, at 20~GHz and on timescales of a few years, a
median debiased variability index, that takes into account the
uncertainties in individual flux density measurements, of 6.9 per
cent, uncorrelated with the flux density, with only a few sources
more variable than 30 per cent. Also, a good fraction (201 sources
corresponding to the 63 per cent of the sample) of our sources are
ATCA calibrators and have therefore been observed repeatedly.
Again, the variability turns out to be relatively modest. Thus,
variability does not affect source counts, since it implies, on
average, an equal number of sources to change to lower or higher
values of flux density. A much better assessment of variability
will be provided by the analysis of the full AT20G data. Since we
selected the observation to which we applied the selection
threshold on the basis of its quality and not on the basis of the
flux density itself (i.e. the best observation is not necessarily
that with the higher value of flux density) we avoid any bias
towards higher flux density values that could affect the source
counts.

\section[]{Properties of the sample}\label{sec:Properties}

\subsection{Radio spectra}

\begin{figure}
\begin{center}
\includegraphics[width=6cm, angle=90]{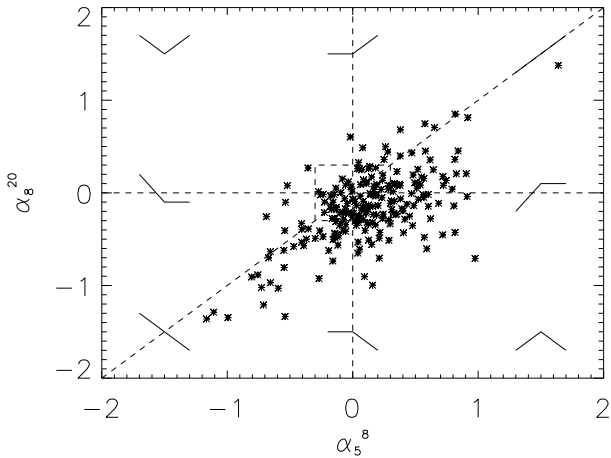}
\caption{Colour-colour radio plot for the 218 sources with near
simultaneous observations: the comparison of the spectral
behaviour in two ranges of frequencies shows the distribution of
the spectral shapes in the whole sample.
Power-law spectra sources lie on the  dashed diagonal line.
A general steepening of the spectra from low (5 to 8~GHz) frequency
to high (8 to 20~GHz) is clearly shown by the large number of sources with
$\alpha_8^{20}<\alpha_5^8$.} \label{fig:cc}
\end{center}
\end{figure}

The spectral index between the frequency $\nu_1$ and $\nu_2$ is
defined as
\begin{equation}
\alpha_1^2 =\frac{\log ({S_1}/{S_2})}{\log( {\nu_1}/{\nu_2})}\
\end{equation}
i.e. $S_\nu\propto\nu^{\alpha}$. Figure~\ref{fig:cc} shows the so
called colour-colour radio plot (Kesteven et al.\ 1977): it is the
comparison of spectral indices at low and high frequencies. Only
the almost simultaneous data have been used in this analysis: the
sub-sample consists of 218 sources. The diagram shows the variety
of spectral behaviours, with a relatively small number of
power-law spectra. Most of the points lie below the diagonal in
Fig.~\ref{fig:cc}, which implies that most sources steepen with
increasing frequency. The median of the difference of the spectral
indices $\alpha_8^{20}-\alpha_5^8$ is $-$0.26 and the standard
deviation of its distribution is 0.34 (see also
Fig.~\ref{fig:steepening_z}). That implies that assuming a simple
power law spectral index equal to $\alpha_5^8$ to extrapolate from
8 to 20~GHz could result, on average, in a 36~per cent error in
the flux density estimation. Thus, simple extrapolations in
frequency using low-frequency spectral indices are highly
unreliable.

\begin{figure*}
\centering
\includegraphics[width=11cm,angle=90]{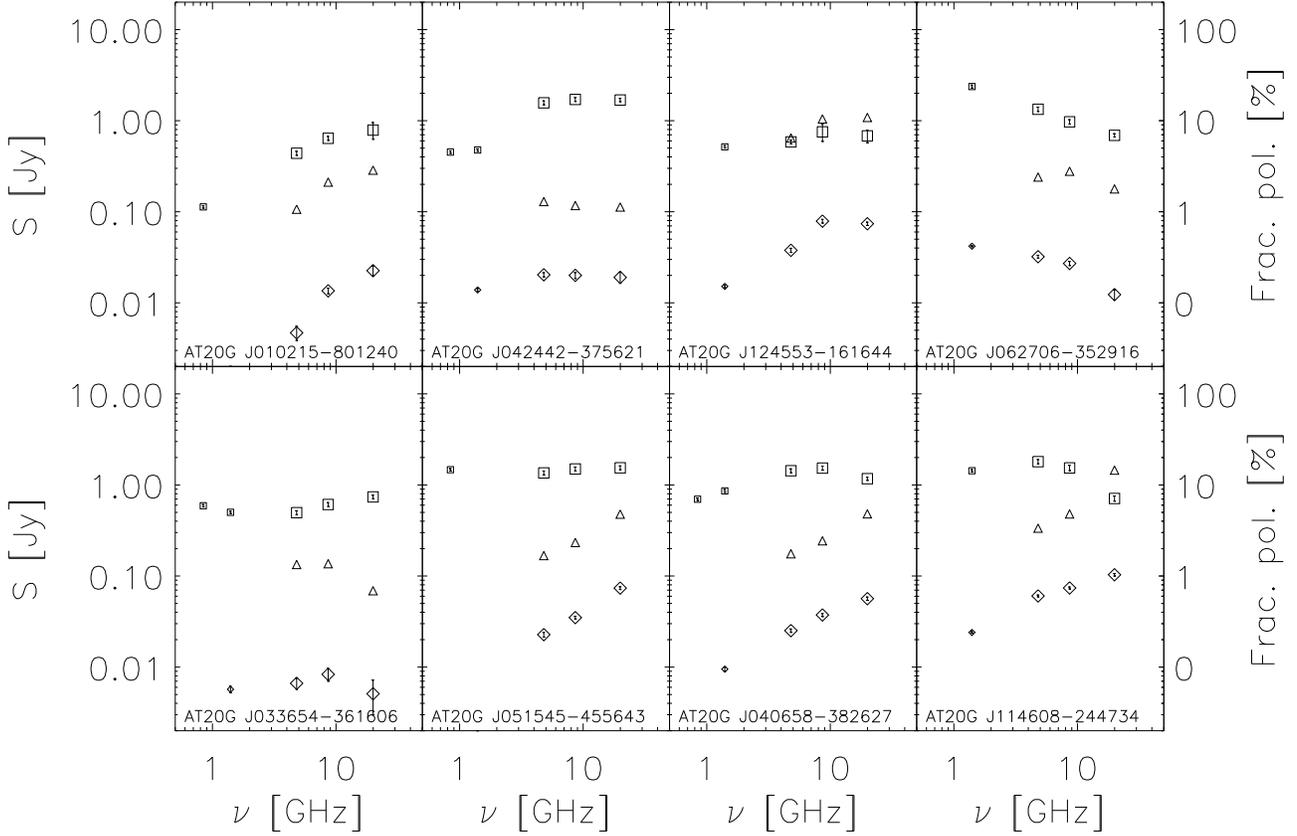}
\vspace*{0.5cm} \caption{Some spectra as example of the large
variety of spectral behaviour in total intensity (squares) and
polarisation (diamonds) for a set of point sources. We selected
examples of inverted, flat, peaked and steep total intensity
behaviour similar (top panels) and different (bottom panels) to
the polarisation behaviour. The triangles correspond to the
fraction of polarisation. The low frequency values refer to data
from SUMSS (0.843~GHz) and NVSS (1.4~GHz) catalogues in total
intensity (small squares) and, where available, polarisation
(small diamonds).}\label{fig:spectra}
\end{figure*}

In Table~\ref{tab:alphatable} we have classified the spectra
shapes on the basis of the spectral indices between 5 and 8~GHz
and between 8 and 20~GHz. Examples of spectra in total intensity
and polarisation are plotted in Fig.~\ref{fig:spectra}, where the
NVSS and SUMSS measurements at 1.4 and 0.843\,GHz are also shown.
Table \ref{tab:alphatable} also gives the fractions of `steep'-
and `flat'-spectrum sources, based on the commonly used
classification (spectral indices smaller or larger than $-0.5$).
We note that in our sample if we separate the compact and extended
sources, the flat (compact) and the steep (extended) populations
overlap for $-0.3<\alpha_8^{20}<-0.5$. This will be discussed in
more details in forthcoming papers on the whole Survey sample.

\begin{figure}
\begin{center}
\includegraphics[width=6cm, angle=90]{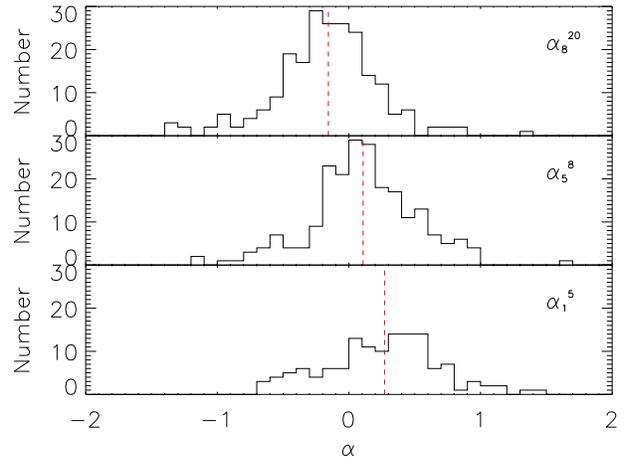}
\vspace*{0.5cm} \caption{Distributions of spectral indices $
\alpha_8^{20}$ (upper panel), $ \alpha_5^8$ (central panel),  and
$ \alpha_{1}^{5}$ (bottom panel). Data at $\sim 1$~GHz come from
NVSS. The red dashed lines correspond to the respective median
values (respectively from the bottom to the top 0.27, 0.11,
$-$0.16).} \label{fig:alphasrccnt}
\end{center}
\end{figure}

\begin{figure}
\begin{center}
\includegraphics[width=6cm, angle=90]{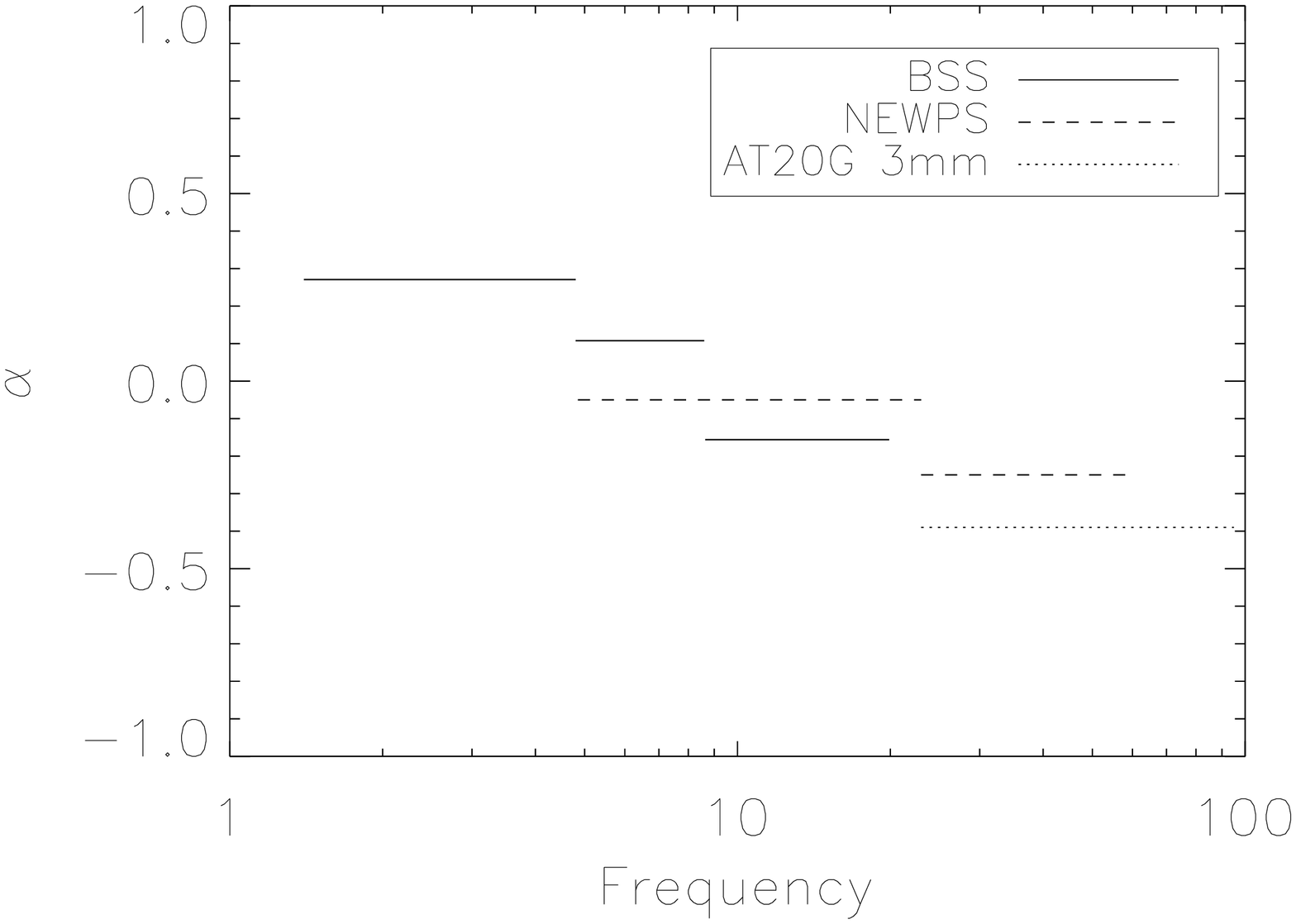}
\caption{Plot of the median spectral indices as they have been
calculated for each frequency range for the BSS (solid lines),
compared with NVSS to get the value between 1.4 and
4.85~GHz, for the NEWPS catalogue (dashed lines) and for the observations
at 95~GHz of a flux limited sample of the AT20G Survey (dotted
line).} \label{fig:steepening}
\end{center}
\end{figure}

\begin{table}
\setcounter{table}{3}
 \caption{Distribution of spectral behaviour for the
  218 BSS sources with almost simultaneous 5, 8 and 20~GHz data.
  The abbreviations in the parenthesis in the second column refer to the classification used to
  flag the sources according to their spectral behaviour in Table 2.
  In the third column there are the numbers of object for each spectral class including a separate
  `very flat' source class. No selection has been applied for flat sources to get the numbers in the last column.
  See the text for details.}
  \label{tab:alphatable}
  \begin{tabular}{llcc} \hline
                  &  & \textbf{No.\tiny{(\%)}}& \textbf{No.\tiny{(\%)}} \\
\textbf{Spectrum} &  & \tiny{\textbf{incl. flat class}}&\tiny{\textbf{excl. flat class}}  \\
 \hline
$\alpha_5^8>0$, $\alpha_8^{20}>0$   & \small{Inverted (I)} & 39 \tiny{(17.9)}& 58 \tiny{(26.6)} \\
\noalign{\smallskip}
$\alpha_5^8>0$, $\alpha_8^{20}<0$   & \small{Peaked (P)}   & 51 \tiny{(23.4)}& 82 \tiny{(37.6)} \\
\noalign{\smallskip}
$\alpha_5^8<0$, $\alpha_8^{20}>0$   & \small{Upturning (U)}& 2 \tiny{(0.9)}  & 9 \tiny{(4.1)}  \\
\noalign{\smallskip}
$\alpha_5^8<0$, $\alpha_8^{20}<0$   & \small{Steep (S) }   & 44 \tiny{(20.2)}& 69 \tiny{(31.7)} \\
\noalign{\smallskip}
$-0.3<\alpha_5^8<0.3$ \&                       &          &           \\
\noalign{\smallskip}
$-0.3<\alpha_8^{20}<0.3$            & \small{`Very' Flat (F)}& 82 \tiny{(37.6)}&         \\
\hline
$\alpha_8^{20}<-0.5$                & \small{Steep }   &  34 \tiny{(15.6)}& \\
\noalign{\smallskip}
$\alpha_8^{20}>-0.5$                & \small{Flat }    &  184 \tiny{(84.4)}& \\
\noalign{\smallskip}
$\alpha_5^8<-0.5$                   & \small{Steep }   &  18   \tiny{(8.3)} & \\
\noalign{\smallskip}
$\alpha_5^8>-0.5$                   & \small{Flat }    &  200  \tiny{(91.7)} &\\
\hline
\end{tabular}
\end{table}
As expected, the 20~GHz sample is dominated by flat-spectrum
sources. A significant trend towards a steepening of spectral
indices at higher frequencies can be noted (see
Fig.~\ref{fig:alphasrccnt}). The median spectral index between 5
and 8~GHz is $0.11$ and the fraction of `steep'-spectrum sources
is $\simeq 8$ per cent, while between 8 and 20~GHz the median
spectral index steepens to $-0.16$ and the fraction of
`steep'-spectrum sources almost doubles to $\simeq 15.5$ per cent.
A similar behaviour has been reported by Bolton et al.\ (2004). It
appears to be more significant at higher frequencies
(Fig.~\ref{fig:steepening}; cf.\ also Gonz{\'a}lez-Nuevo et al.\
2007). An even larger steepening effect was found for a deeper
($S_{\rm lim, 20 GHz}> 150$\,mJy) selected sample of the AT20G
Survey (Sadler et al.\ 2007).
The presence of spectral curvature provides valuable information
about the physical conditions in a radio source. Two mechanisms
which generate spectral curvature are the energy losses by
synchrotron radiation causing steepening of the spectrum at high
frequencies (e.g., Pacholczyk\ 1970) and optical depth effects in
compact sources at lower frequencies which may be due to either
free-free opacity or synchrotron self absorption. The clear
evidence for spectral steepening of integrated flux density in the
majority of the sources in the 20\ GHz Bright Source Sample
(Fig~\ref{fig:cc}) and the increased spectral steepening observed
at higher frequencies (Fig.~\ref{fig:steepening}) is in stark
contrast to the lack of spectral steepening in the integrated flux
density for radio sources in low frequency surveys (e.g. Laing and
Peacock\ 1980). Spectral steepening in the resolved structure in
radio source lobes is commonly seen and successfully modelled by a
combination of energy losses and continual reacceleration in the
lobes (e.g., Jaffe \& Perola\ 1973, Subrahmanyan et al.\ 2006).

The class of flat and inverted spectrum objects which
dominates the high frequency AT20G sample is quite different.
The objects are small and almost certainly in a younger evolutionary phase
which includes the `Gigahertz Peaked Spectrum' (GPS) sources
(e.g., O'Dea\ 1998, Tinti \& De Zotti\ 2006).

The spectral steepening of sources in the lower left quadrant in
Fig.~\ref{fig:cc} could be due to synchrotron aging which would be
much more rapid in the compact radio sources because the magnetic
fields are higher.

The BSS sample contains 64 objects (29.4
per cent of the 218 objects with simultaneous observations at 5, 8
and 20~GHz) with $\alpha_5^8>\alpha_8^{20}$ and $\alpha_5^8>0.3$,
i.e. peaking above 5~GHz.
Tinti et al.\ (2005) argued that a large fraction of sources
showing spectral peaks at several GHz are not truly young (GPS)
sources but blazars where a flaring, strongly self-absorbed
synchrotron component, probably originated at the base of the
relativistic jet, transiently dominates the emission spectrum.
Although our evidence for variability (see \S~\ref{sec:completeness}) does not
provide much support for this model, repeated simultaneous multifrequency measurements with time lags
of a few years will be needed to discriminate among the two populations.
Polarisation measurements are also a good discriminant, as true
GPS sources generally have much lower polarisation levels than
blazars  (Orienti \& Dallacasa\ 2007). In Fig.~\ref{fig:mS20} we have separated
the sources with peaks in the spectrum above 5~GHz. The most unambiguous
discrimination is however obtained with high resolution radio
interferometry, observing the different milli-arcsec morphology of
blazars and GPS sources.

\subsection{Extended sources}\label{sec:extended}
\begin{figure}
\begin{center}
\includegraphics[width=6cm, angle=90]{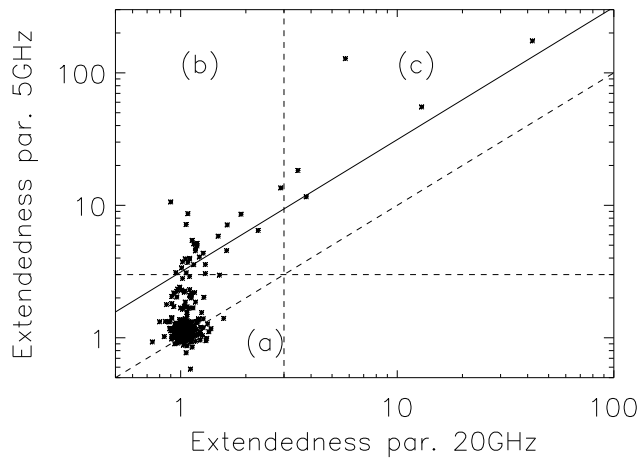}
\caption{5~GHz versus 20~GHz extendedness parameter (see the text
for details).} \label{fig:extpar}
\end{center}
\end{figure}

\begin{table*}
 \caption{Table of extended sources in the BSS. The first column lists the sequential number of the sources as in Table~2.
 An `M' indicates that they have been observed in mosaic mode.
 The 20~GHz flux densities in column 4 refer to the core region whereas those in column 5 are the integrated flux densities.
 For 3 sources observed in mosaic mode, we believe we have acquired the flux density values only for subregions,
 so we consider them as lower limits of the total integrated flux densities.
 P.A. is the position angle (in degrees) of the major axis of the source.}
  \label{tab:exttable}
  \begin{minipage}{190mm}
  \begin{tabular}{llllllllllll}
\hline
Seq.&RA&$\delta$&$S_{20\,GHz}$&$S_{20\,GHz}$&$S_{8.6\,GHz}$&$S_{4.8\,GHz}$&$P_{20\,GHz}$&z &Size&P.A.&\tiny{Alternative}\\
 \# & &  & $core[Jy]$ & $[Jy]$& $[Jy]$ & $[Jy]$ & $[Jy]$ & & \tiny{[arcmin]} & [$^\circ$]& \tiny{Name}  \\
\hline
~20M&01:33:57.6 &-36:29:34.9&0.041\tiny{ 0.005} &$>$1.86\tiny{   ..}&... \tiny{   ..}&... \tiny{   ..}&  ...    &  ...  & 6.1& 79 &\tiny{PKS 0131-36}\\
~52 &04:08:48.75&-75:07:20.1&  ...              &0.86\tiny{ 0.14}   &2.64\tiny{ 0.42}&4.74\tiny{ 0.75}&  ...    &0.693  & 0.1 & 45  &\tiny{PKS 0410-75}\\
~69M&05:19:49.7 &-45:46:44.2&1.33\tiny{ 0.07}   &8.52\tiny{ 0.11}   &... \tiny{   ..}&... \tiny{   ..}&  1.400\tiny{ 0.016}&0.0351& 3.4& 76 &\tiny{Pictor A}\\
~71 &05:22:57.94&-36:27:30.4&  ...              &3.91\tiny{ 0.59}   &6.57\tiny{ 1.04}&9.07\tiny{ 1.43}&  ...    &0.0553 & 0.5& 55 &\tiny{PKS 0521-36}\\
~92 &06:35:46.33&-75:16:16.9&  ...              &3.24\tiny{ 0.51}   &4.82\tiny{ 0.76}&5.54\tiny{ 0.87}&  ...    &0.653  & 0.2& 90 &\tiny{PKS 0637-75}\\
100 &07:43:31.60&-67:26:25.7&  ...              &1.22\tiny{ 0.19}   &1.87\tiny{ 0.30}&2.34\tiny{ 0.37}&  ...    &1.51& 0.2& 14 &\tiny{PKS 0743-67}\\
118 &09:19:44.06&-53:40:05.1&  ...              &0.94\tiny{ 0.15}   &1.69\tiny{ 0.27}&2.5 \tiny{ 0.39}&  ...    &  ...  & 0.3  & 40 &\tiny{PMN J0919-5340}\\
157 &12:05:33.37&-26:34:04.9&  ...              &0.84\tiny{ 0.14}   &1.18\tiny{ 0.21}&1.12\tiny{ 0.18}&  ...    &0.789  & 0.4  & .  &\tiny{PKS 1203-26}\\
179M&13:25:27.7 &-43:01:07.0&7.62\tiny{ 0.44}   &$>$59.3\tiny{   ..}&... \tiny{   ..}&... \tiny{   ..}&  ...    &0.00183&10.9& 34 &\tiny{Centaurus A}\\
182M&13:36:39.0 &-33:57:58.2&0.21\tiny{ 0.04}   &$>$1.60\tiny{   ..}&... \tiny{   ..}&... \tiny{   ..}&  ...    &0.01254&31  & 53 &\tiny{PKS 1333-33}\\
185 &13:46:48.95&-60:24:29.0&  ...              &5.30\tiny{ 0.84}   &6.14\tiny{ 0.97}&6.58\tiny{ 1.04}&  ...    &  ...  & 0.1  & .  &\tiny{PKS 1343-60}\\
216M&16:15:05.2 &-60:54:25.5&0.19\tiny{ 0.05}   &3.84\tiny{ 0.04}   &... \tiny{   ..}&... \tiny{   ..}&0.169\tiny{ 0.011}&0.01828&12.6& 47 &\tiny{PKS 1610-60}\\
284M&21:57:06.08&-69:41:23.3&  ...              &5.31\tiny{ 0.27}   &... \tiny{   ..}&... \tiny{   ..}&0.087\tiny{ 0.014}&0.0283 & 1.2& 20 &\tiny{PKS 2153-69}\\
310 &23:33:55.28&-23:43:40.8&  ...              &0.82\tiny{ 0.13}   &1.47\tiny{ 0.22}&0.67\tiny{ 0.10}& ...     &0.0477 & 21 & -43 &\tiny{PKS 2331-240}\\
319M&23:59:04.70&-60:55:01.1&0.11\tiny{ 0.06}   &3.03\tiny{ 0.05}   &... \tiny{   ..}&... \tiny{   ..}&0.053\tiny{ 0.008}&0.0963 & 6.3& 46 &\tiny{PKS 2356-61}\\
\hline
\end{tabular}
\end{minipage}
{Note: sources number 20, 69, 182, 284, 310 and 319 are
characterized by a core and double lobes; 71, 92 and 100 have a
core and a jet; 179 is the inner double lobe of the giant radio
galaxy Centaurus A with total extent of 5 degrees; 216 is a wide
angle tail source; 310 is the core region of a highly-extended
radio galaxy: it is difficult to determine the correct size
without a mosaic observation. References for redshift are given in
Table~2. Useful references for the single sources are as follows.
20: Ekers et al.\ (1978); 69: Perley, Roser, \& Meisenheimer\
(1997); 71: Birkinshaw, Worrall, \& Hardcastle\ (2002);92:
Schwartz et al.\ (2000); 182: Killeen, Bicknell, \& Ekers\ (1986);
284: Fosbury et al.\ (1998).}
\end{table*}
The comparison of the extendedness parameters at different
frequencies (Fig.~\ref{fig:extpar}) for the BSS sources confirms
the expectation that the extended, steep-spectrum radio lobes are
less and less prominent at higher frequencies.

\begin{figure}
\begin{center}
\includegraphics[width=6cm, angle=90]{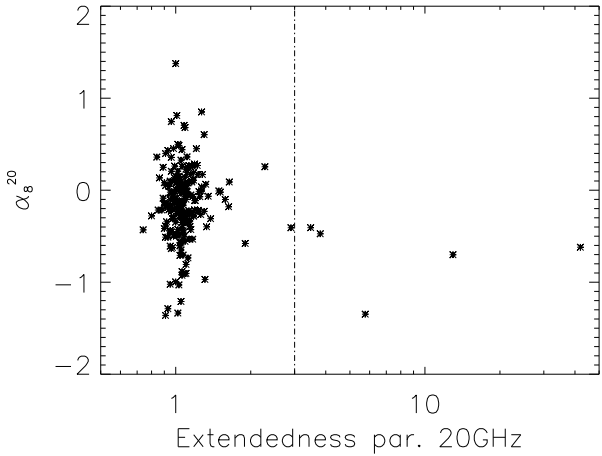}
\caption{The spectral indices between 8 and 20~GHz versus the
20~GHz extendedness parameter.} \label{fig:alpha_ext}
\end{center}
\end{figure}

In Fig.~\ref{fig:extpar} we can see three clear effects. There are
point sources spread into a circular patch by noise (a), a group
of sources extended at 5~GHz but still dominated by a point core
at 20~GHz (b) and a group of sources extended at both 5 and 20~GHz
(c) which have a steeper spectral index for the extended
component. The solid line corresponds to $\alpha_8^{20}=-0.8$
which is the expected spectral index for extended components of
radio galaxies and QSO (cf. Laing \& Peacock\ 1980). The median
8--20~GHz spectral index of the extended objects ($-$0.62) is
similar to sources found in low frequency samples
(Fig.~\ref{fig:alpha_ext}). In general, 8 of the 39 extended
objects at 5~GHz are extended also at 20~GHz (considering also 3
non-simultaneous cases that do not appear in Fig.~\ref{fig:extpar}
and \ref{fig:alpha_ext}). To those we could add 7 extended sources
at 20~GHz, for which we don't have low frequency data, but we know
from other catalogues that they are extended also at low
frequencies.

As anticipated in \S\,\ref{sec:completeness} we have looked for
extended sources missed by the BSS selection because they are
either: 1) fully resolved (and therefore undetected) by the 60~m
shortest antenna spacings used in the follow-up, or 2) had
components (hot-spots, cores) which have been detected as
individual sources in the AT20G follow-up.  An inspection of the
SUMSS (Mauch et al.\ 2003) and of the PMN (Griffith \& Wright\
1993) catalogues yielded 9 sources that are extended, bright, and
with 0.84--5~GHz spectral indices such that the expected integral
flux densities at 20~GHz may be $> 0.50$~Jy, (that happens in 7 cases
that are flagged with an `M' in Table~2 and in
Table~\ref{tab:exttable}) but present in the initial BSS selection
only if their core component has flux density above $0.50\,$Jy
(see Table~\ref{tab:exttable}). All of these have been observed
with the mosaic mode. For these sources we have integrated flux
densities at 20~GHz but no flux densities at lower frequencies.
Therefore we could not determine the extendedness parameter at low
frequencies or the spectral indices for them, that are thus
missing in Fig.~\ref{fig:extpar} and \ref{fig:alpha_ext}.

A summary of the properties of the extended sources in the BSS is
in Table~\ref{tab:exttable}. A few more sources  that lie at the
edge of our classification have been listed and commented in the
Appendix~\ref{sec:IndSrc}.

\subsection{Polarisation}
\begin{table}
 \caption{Matrix of the number of objects according to the combination of the total intensity and polarisation spectral behaviour for the sources with almost simultaneous total intensity and polarisation detection at 5,8 and 20~GHz.
 On the rows there are the spectra shapes in polarisation, on the columns the spectra shapes in total intensity.
 The spectral types are defined in Table~\ref{tab:alphatable}.} \label{tab:matrix}
  \begin{tabular}{|l|c|c|c|c|c|}
  \hline
  \textbf{$S$ $\rightarrow$}&\textbf{U}&\textbf{I}&\textbf{F}&\textbf{P}&\textbf{S}\\
  \textbf{$Pol.$ $\downarrow$ } &&&&&\\
  \hline
  \textbf{U}    & 0  &  2   &  7 &  1 & 2 \\
  \textbf{I}     & 1  &  7  & 16 & 10 & 3 \\
  \textbf{F}         & 0  &   0  &  5 &  1 & 2 \\
  \textbf{P}       & 0  &   7  & 24 & 13 & 5 \\
  \textbf{S}        & 0  &   1  &  6 &  2 & 8\\
  \hline
\end{tabular}
\end{table}
\begin{table}
 \caption{The same as Table \ref{tab:matrix}, but on the rows there are the spectral shape of the fractional polarisation.
 The spectral types are defined in Table~\ref{tab:alphatable}.} \label{tab:matrix_m}
  \begin{tabular}{|l|c|c|c|c|c|}
  \hline
  \textbf{$S$ $\rightarrow$}&\textbf{U}&\textbf{I}&\textbf{F}&\textbf{P}&\textbf{S}\\
  \textbf{$m[\%]$ $\downarrow$ } &&&&&\\
  \hline
  \textbf{U}    & 0 &  5   &  9 &  5 & 7 \\
  \textbf{I}     & 1  &  4  & 14 & 13 & 4 \\
  \textbf{F}         & 0  &   0  &  6 &  2 & 3 \\
  \textbf{P}       & 0  &   4  & 18 & 2 & 3 \\
  \textbf{S}        & 0  &   4  &  11 &  5 & 3 \\
  \hline
\end{tabular}
\end{table}
All the follow-up measurements include polarisation. Once the low
quality data have been removed from the sample, we take, as
`detections', measurements of integrated polarised flux at least 3
times higher than their errors (see \S~\ref{sec:polarisation}).

We had a polarisation detection at 20~GHz for 213 sources (34
cases are non detections, the others have low quality data in
polarisation and the data have not been considered). The median
fractional polarisation is 2.5 per cent, calculated considering
also upper limits with a Survival Analysis procedure. The median
polarisation degree is found to be somewhat lower at lower
frequencies: it is 2.0 per cent at 8~GHz and 1.7 per cent at 5~GHz
(see Fig.~\ref{fig:mS}). A similar trend was found by Burke et
al.\ (in prep.) for the sub-sample observed in October 2006 during
the observation run dedicated to high sensitivity polarisation
observations. A detailed analysis of polarisation data will be
presented in that paper.

As can be seen from Fig.~\ref{fig:spectra} the spectra for polarised
flux density are very diverse and show little correlation with
total flux density. This makes it even more difficult to predict
high frequency polarisation properties from low frequencies
observations than it is to predict $I$.

There is no clear relation between the spectral properties of the
sources and their polarised flux, nor is there any unique trend in
the spectral behaviour of the total intensity and the polarised
emission.  The spectral shape in the polarisation is often quite
different from the spectral shape in the total intensity.

The matrices of spectra in Tables~\ref{tab:matrix} and ~\ref{tab:matrix_m} are complex but not
random. The diagonal cells dominate indicating that nearly 50 per cent
of the sources have polarised spectra similar to those in $I$.
However the flat and peaked spectrum sources stand out with an
excess of rising polarisation spectra.

For sources with peaked spectra the polarised fraction generally
decreases below the turnover frequency; an example of this
behaviour in Fig.~\ref{fig:spectra} (third panel, bottom row).
This is not surprising as a polarisation mode which has a high
emission coefficient should also have a high absorption
coefficient, so in moving from optically thin (high frequencies)
to optically thick (low frequencies) conditions we expect that the
ratio of the intensities in the two modes will decrease. There
are, however, other reasons why the polarised fraction might
decrease at lower frequencies, including:
\begin{itemize}
\item depolarisation due to Faraday rotation intrinsic to the
sources; \item superposition of multiple components with different
polarised spectra; \item depolarisation due to spatial variations
in Faraday Rotation across the source; \item bandwidth
depolarisation due to very high levels of Faraday Rotation.
\end{itemize}

\begin{figure}
\begin{center}
\includegraphics[width=6cm, angle=90]{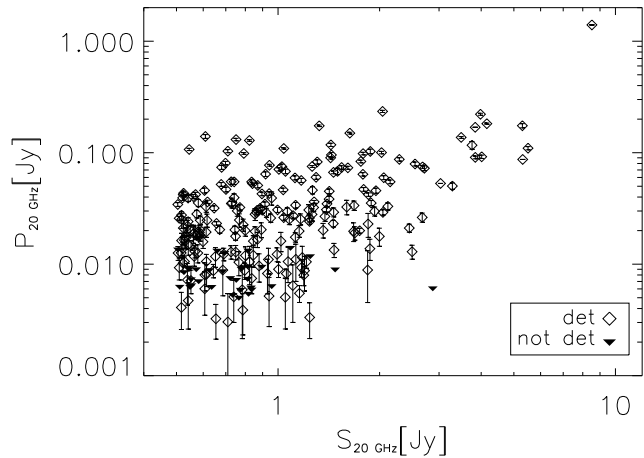}
\caption{Integrated polarised flux as a function of total
intensity 20~GHz flux. The bright source at $P=1.4$\,Jy is
Pictor~A.} \label{fig:PS}
\end{center}
\end{figure}

\begin{figure}
\begin{center}
\includegraphics[width=6cm, angle=90]{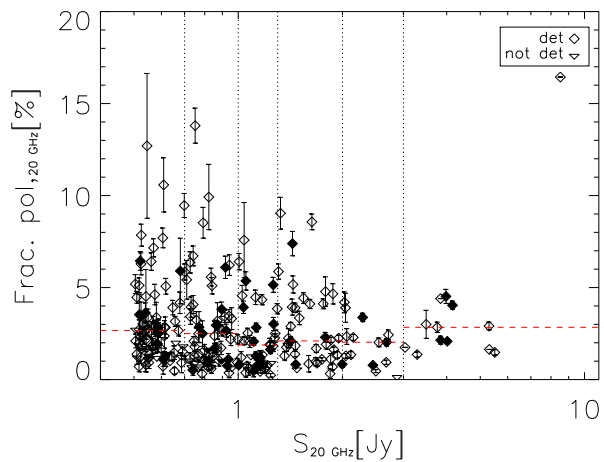}
\caption{Fractional polarisation as a function of total intensity
20~GHz flux density. The dashed lines shows the median fractional
polarization by bins (the dotted lines indicates the bin ranges)
of flux density for the full sample. Filled simbols refer to
objects with $\alpha_5^8>\alpha_8^{20}$ and $\alpha_5^8>0.3$.}
\label{fig:mS20}
\end{center}
\end{figure}

\begin{figure}
\begin{center}
\includegraphics[width=6cm, angle=90]{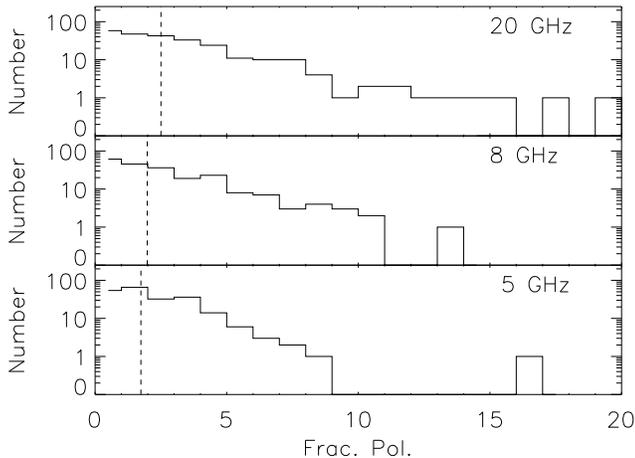}
\caption{Distribution of fractional polarisation at 5, 8 and
20~GHz. Dashed lines are the median values.} \label{fig:mS}
\end{center}
\end{figure}

Figures~\ref{fig:PS} and \ref{fig:mS20} plot the polarised flux
and fractional polarization as a function of flux density.  The
data from our pilot observations (Sadler et al 2006) suggested a
marginal trend for weaker sources to have higher fractional
polarization.  Although this seems to be present in
Figure~\ref{fig:mS20} the median fractional polarization as a
function of flux density has no trend and indicates that the
apparent effect is due to the increased density of points at lower
flux levels. The sources with a peak in the spectrum above 5~GHz
have lower fractional polarization at 20~GHz  but this effect is
not very pronounced. Fig.~\ref{fig:mS} shows the distribution of
fractional polarisation at 5, 8 and 20~GHz.
\subsection{Radio counterparts and flux density comparisons}

\subsubsection{Low frequency catalogues}

Due to the lack of deep large area surveys at frequencies above
15\,GHz the comparison of our results has to be done with low
frequency catalogues. Because of variability between catalogue
epochs a direct comparison can only provide hints on the spectral
behaviour as discussed in the previous section. The results of the
cross-correlation with NVSS at 1.4~GHz and SUMSS at 0.843~GHz have
been listed in Table 2. All 172 BSS sources in the sky region
overlapping with the NVSS survey have at least one counterpart in
NVSS (within less than 1.2~arcmin from the position of the BSS
source). A total of 149 BSS sources have a counterpart in SUMSS.

\begin{figure}
\begin{center}
\includegraphics[width=6cm, angle=90]{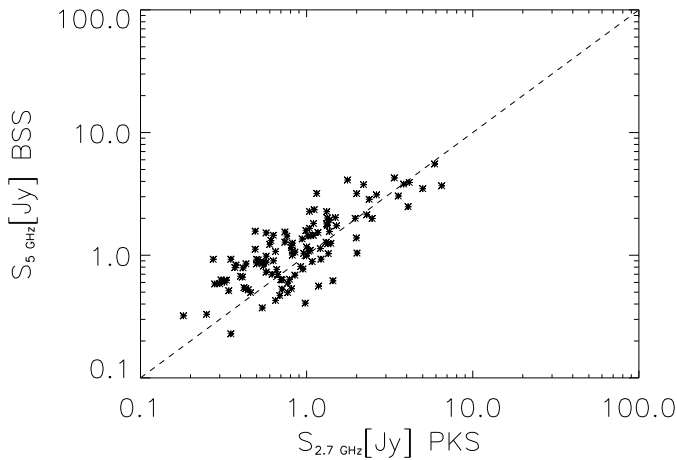}
\caption{Comparison of 5\,GHz flux densities with the Parkes
quarter Jy sample.} \label{fig:PKS}
\end{center}
\end{figure}

\begin{figure}
\begin{center}
\includegraphics[width=6cm, angle=90]{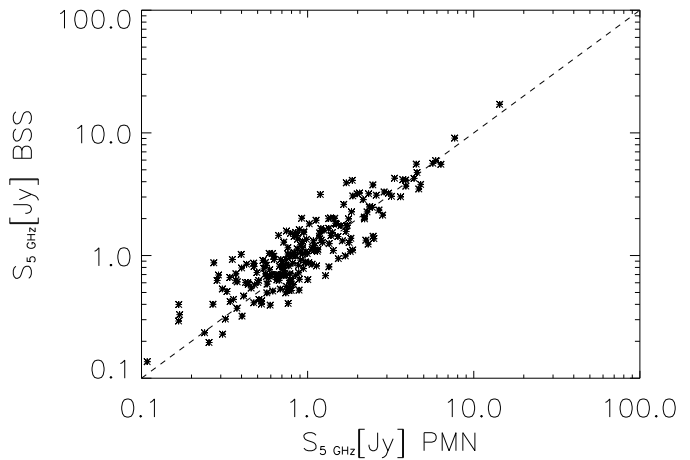}
\caption{Comparison of 5\,GHz flux densities with the PMN
catalogue.} \label{fig:PMN}
\end{center}
\end{figure}

At 2.7~GHz we have cross-matched the BSS with the Parkes quarter
Jy sample (Jackson et al.\ 2002). Of the 314 BSS sources in the
overlapping declination range, 163 have a counterpart. At 4.85~GHz
the cross-correlation with the PMN catalogue shows that 316 BSS
sources have a counterpart in PMN.  The four BSS sources without a
PMN counterpart lie in the small regions of sky where the PMN
survey was incomplete (see, e.g., Figure~2 of Wright et al.\
1996). The 4.8~GHz flux densities from our observations have been
used for comparison with these two catalogues (see
Fig.~\ref{fig:PKS} and \ref{fig:PMN}). The closeness in frequency
reduces the spectral effects and the scatter mainly results from
variability. The few sources that fall below $\sim0.4$~mJy have
the most inverted spectra since our sample is flux limited at
20~GHz.

There are 88 BSS sources in the 185 sources monitored with the
ATCA at 1.4, 2.5, 4.8 and 8.4~GHz at up to 16 epochs by Tingay et
al.\ (2003). In addition to fractional polarisations at each
frequency, and a measure of source extendedness, the multi-epoch
monitoring enabled a variability index to be assigned for each
frequency. The monitoring was done to support  the VSOP Survey
Program, and 87 BSS sources are included in the 5~GHz survey of
bright compact AGN (Hirabayashi et al.\ 2000). Results from the
these space VLBI observations are presented by Scott et al.\
(2004) and Dodson et al.\ (2007), and will be discussed in more
detail in a later paper describing the VLBI properties of the BSS.

\subsubsection{Interest for CMB missions}
The contamination due to point sources is a crucial limitation to
the CMB power spectrum determination on the smaller angular scales
(less than $\sim 30$~arcmin).

The best frequency region to study the CMB is around 70~GHz, where
the effects of foregrounds emissions is at a minimum, but in any
case it is necessary to enlarge the frequency range as much as
possible to try to single out all the foreground components and
improve the component separation techniques.  However, the
efficiency of these methods relies on a good knowledge of the
source populations to improve the capabilities of blind detection
methods for `point' sources (L{\'o}pez-Caniego et al.\ 2007).

The variety of source spectral behaviours implies that, as
mentioned, it is extremely difficult to make reliable flux density
extrapolations from low to high frequency. Variability and
confusion effects complicate the situation even more.

Also, the forthcoming Planck  mission will be strongly confusion
limited. According to L{\'o}pez-Caniego et al.\ (2006), the
$5\sigma$ detection limits range from $\simeq 520\,$mJy at 30~GHz
to $\simeq 180\,$mJy at 100~GHz, while the rms noise levels are
far lower (from $\simeq 19\,$mJy at 30~GHz , Valenziano et al.\
2007, to $\simeq 14\,$mJy at 100~GHz; Lamarre et al.\ 2003): this
means  that there is a lot of astrophysical information in Planck
maps below the  confusion limit, that can be to some extent
extracted, e.g. using stacking  techniques, thanks to the AT20G
Survey and follow-up observations at higher frequencies.

As a test of high frequency predictions from low frequency samples
we selected a sample from the PMN catalogue with declination
below $-30^\circ$ and $|b|>10^\circ$ and cross-matched it with SUMSS
to obtain the low frequency spectral behaviour. Then we divided
it into sub-samples according to different limits in flux density
at 5 and 1\,GHz and/or according to different spectral indices at
those frequencies. Finally, for each sub-sample we considered how
many PMN sources are in the sample and how many of them have a
counterpart in the BSS. In fact, the efficiency of the detection
depends on the ratio between what is present at the selection
frequency and what is effectively found at the detection frequency
(\emph{detection rate}), and on the completeness of the sample
obtained at the detection frequency.

There are 154 BSS sources with declination below $-30^\circ$ and
$|b|>10^\circ$ and 152 have a PMN counterpart. However, 35 PMN
counterparts have flux density at 5\,GHz below 0.50\,Jy, so that a
low frequency selection threshold at 500\,mJy would have lost
them. Selecting only inverted sources ($\alpha_{0.843}^{4.85}>0$,
$\alpha_{0.843}^{4.85}>0.25$, $\alpha_{0.843}^{4.85}>0.5$) results
in a low detection rate ($3.6$, $3.2$, $3.0$ per cent
respectively) and a low completeness of the sample at high
frequency ($52.2$, $22.2$, $9.8$ per cent respectively). A flux
density selection at 5\,GHz implies a decreasing completeness with
increasing 5\,GHz flux density threshold (from 90.2 to 71.9 per
cent changing from 250 to 500~mJy) with low, but increasing
detection rate (from 11.3 to 27.5 per cent in the same flux
density range): 289 PMN sources with a counterpart in SUMSS with
declination below $-$30$^\circ$ and $|b|>10^\circ$ have flux
density above 500\,mJy and no counterpart in the BSS. Combining
spectral and flux density limits or adding further selection
criteria at 1\,GHz improves the detection rate but at the cost of
a very low completeness of the high frequency sample. Thus, it is
clear that low frequency catalogues could provide positions
for constrained search techniques (cf.\ L{\'o}pez-Caniego et al.\
2007), but are inadequate to forecast the high frequency
population.

\begin{figure}
\begin{center}
\includegraphics[width=6cm, angle=90]{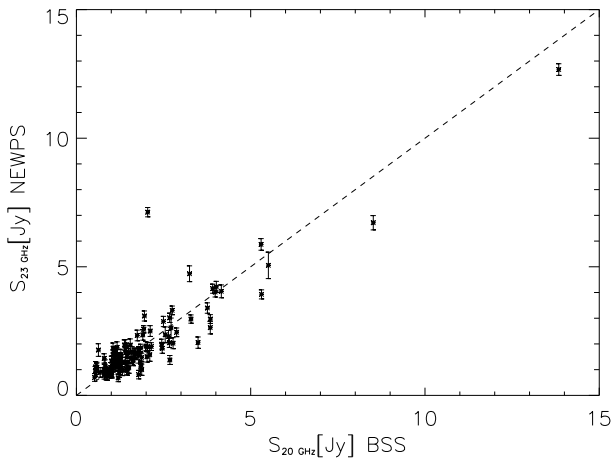}
\caption{Comparison of the BSS 20\,GHz flux densities with the NEWPS
catalogue at 23~GHz.} \label{fig:NEWPS}
\end{center}
\end{figure}

The comparison of flux densities with WMAP map-based catalogues
shows a good agreement in general (we used the NEWPS catalogue as
in Gonz{\'a}lez-Nuevo et al.\ 2007 in Fig.~\ref{fig:NEWPS}). The epochs of observations
partially overlap, but since the WMAP maps have been averaged over
three years, transient phenomena have been smoothed out.

Furthermore, both space- and ground-based missions require a set
of carefully selected sources to work as calibrators for pointing,
total intensity flux densities and polarisation angle.

The Bright Source Sample we have discussed is well-suited for CMB
studies in the next years. It collects a sample of the brightest
sources in the Southern sky, that, thanks to the low variability
observed, will be observable or detectable by any detection
method. The observational frequency is, so far, the closest to the
region of the spectrum of interest for CMB studies.

The Bright Source Sample provides a direct test of source
detection algorithms, quantifying the completeness, the fraction
of spurious detections, the effective beam size (and therefore the
flux calibration) and the possible presence of biases in flux
density estimates. It also provides a rich list of candidate flux
density and pointing calibrators over a large fraction (37 per
cent) of the sky.

Finding suitable polarisation calibrators for CMB experiments is
much more complicate. For example, the large low frequency beams
of the Planck satellite (33 arcmin beam at 30\,GHz) dilute the
polarised signals by summing over differently oriented
polarisation vectors. Thus, finding sources with large enough
polarised flux density within such beams is very hard
(Figs.~\ref{fig:PS} and \ref{fig:mS20}). A more extensive discussion
about such calibration will appear in Burke et al.\ (in preparation).

\subsection{Optical identifications and redshifts}\label{sec:optid}

\subsubsection{Optical identifications}

To make optical identifications for objects in the Bright Source
Sample, we searched the SuperCOSMOS catalogue (Hambly et al.\
2001) near the positions of all sources.
Objects within 10$^\circ$ of the Galactic plane (flagged with a `G' in Table~2) were excluded from the analysis
because the presence of foreground stars and Galactic dust
extinction makes optical identifications incomplete in this
region. This cutoff in Galactic latitude excluded 69 of the 320
BSS sources. Two other sources were also excluded from the
optical analysis: sources number 57 and 160 (according to the sequential
numeration in Table~2) lie so close to bright foreground stars that no optical identification is possible
from the DSS images. Source number 73 lies within the boundaries of the Large
Magellanic Cloud and its identification is uncertain.

An optical object was
accepted as the correct ID if it was brighter than B$_{\rm
J}$=22\,mag and lay within 2.5\,arcsec of the radio position.
Monte Carlo tests imply that at least 97 per cent of such objects are
likely to be genuine associations (Sadler et al.\ 2006).

\begin{figure}
\begin{center}
\includegraphics[width=6cm, angle=90]{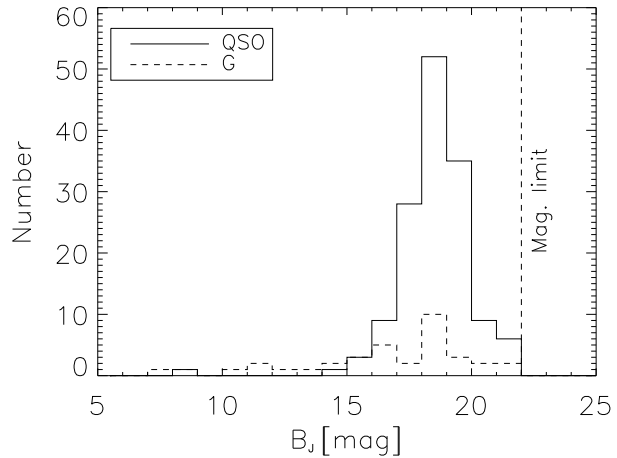}
\caption{B-magnitude distribution.} \label{fig:Bsrccnt4}
\end{center}
\end{figure}

We found a DSS identification for 238 of 249 sources, with 235 of
the optical IDs having $B_{\rm J}\leq22.0$\,mag. On the basis of
the SuperCOSMOS classification of each object as stellar or
extended, there are 188 QSOs (77 per cent of the sample), 47
galaxies (19 per cent) and the remaining are blank fields.
The median $B_{\rm J}$ magnitude is 18.6 for QSOs and 17.7 for
galaxies (see Fig.~\ref{fig:Bsrccnt4}).

We have also checked in the NASA Extragalactic Database
(NED\footnote{http://nedwww.ipac.caltech.edu/}) for optical
identifications in order to distinguish between Galactic and
extragalactic objects: none of the sources in the BSS which have a
clear identification are Galactic objects (i.e. HII regions,
planetary nebulae or SNRs).

\subsubsection{Published redshifts}
After completing the optical identifications, we checked (NED) to
search for published redshifts. A listed redshift was accepted
only if it could be traced back to its original source and
appeared to be reliable. 177 of the 249 BSS objects (71 per cent)
had a reliable published redshift, including three of the sources
which are blank fields on the DSS (these objects were identified
in deeper optical images by other authors).

The 72 objects without a published redshift include seven objects
(sources number 10, 19, 30, 42, 85, 221 and 278 as listed in
Table~2) which have a redshift listed in NED. In these cases, we
were either unable to trace back to its original source, or
considered to be unreliable for other reasons (PKS~0332$-$403,
source 42 in Table~2, was previously discussed in this regard by
Shen et al.\,1998).

\subsubsection{New redshifts}
Redshifts for two BSS objects (sources number 33, a QSO at
$z=$0.466 QSO,  and 313, a QSO at $z=$0.626 based on a single
broad emission line identified as MgII) were obtained from a
pre-release version of the final redshift catalogue from the 6dF
Galaxy Survey (Jones et al.\ 2004, 2007 in preparation). The
redshift for source 138 has been measured with the ESO 3.6 m
telescope by PGE and his collaborators (Edwards et al. in
preparation).

Optical spectra of nine other BSS sources were obtained at the ANU
2.3 m telescope in April and June 2007 by R.W.\ Hunstead and two
of the authors (PH and EM). Redshifts were measured for seven of
these objects (sources number 77, 98, 140, 162, 166, 208 and 246).
The spectra of two other objects (number 68 and 78) showed a
featureless optical continuum from which no redshift could be
measured.

\begin{figure}
\begin{center}
\includegraphics[width=6cm, angle=90]{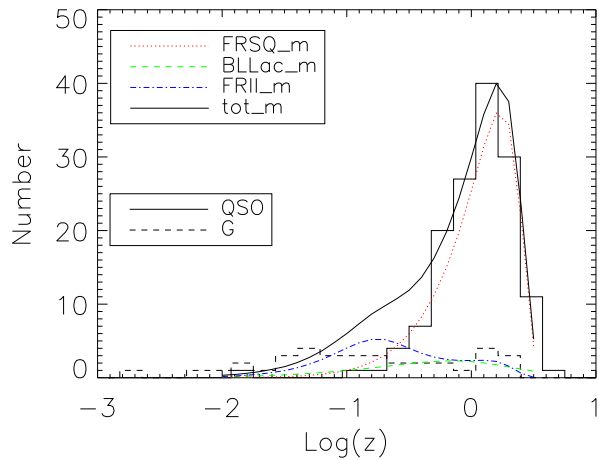}
\caption{Redshift distribution. The model by
De Zotti et al.\ (2005) has been overlapped for comparison.} \label{fig:zsrccnt}
\end{center}
\end{figure}

\begin{figure}
\begin{center}
\includegraphics[width=6cm, angle=90]{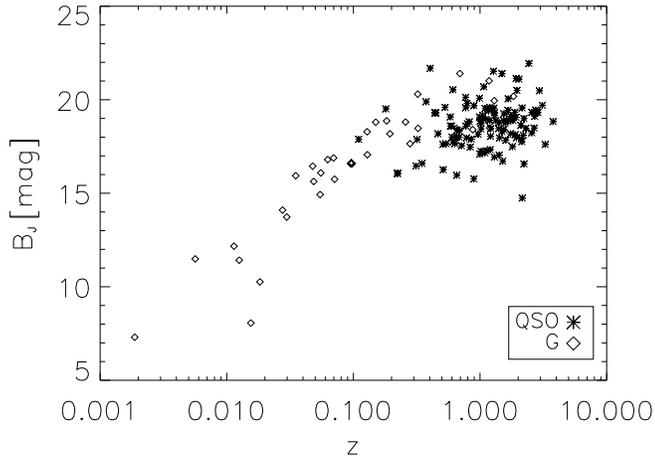}
\caption{B-magnitude versus redshift for galaxies
and QSO.} \label{fig:Bz}
\end{center}
\end{figure}

Among the 186 objects with redshifts, 144 are QSOs and 36 are
galaxies. The median redshift is 1.20 for the QSOs and 0.13 for
the galaxies(Fig.~\ref{fig:zsrccnt}). No correlation is observed
between redshift and total 20~GHz flux density or polarised flux.
As noted by Sadler et al.\ (2006) there is a correlation between
redshift and optical magnitude for galaxies in the AT20G sample,
but this does not apply to the AT20G quasars (see
Fig.~\ref{fig:Bz}).

\subsubsection{Objects with featureless optical spectra}
Six BSS objects with good--quality optical spectra (either from
the published literature or from unpublished 6dF/2.3~m data), have
no measured redshift because the spectra are featureless.  Such
objects generally fall into the BL Lac class, though it is
possible that some of them fall in the `redshift desert' at
$z\sim1.5-2.2$ where QSOs show no strong lines in the optical.

\subsubsection{Spectral properties and redshift}
\begin{figure}
\begin{center}
\includegraphics[width=6cm, angle=90]{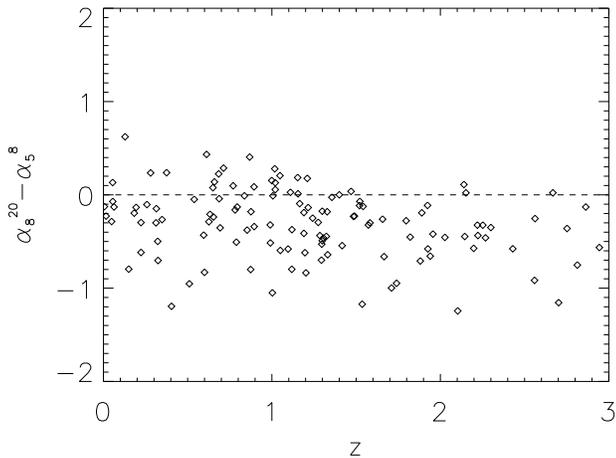}
\caption{Plot of the difference between spectral indices
$\alpha_8^{20}$ and $\alpha_5^{8}$ with redshift.}
\label{fig:steepening_z}
\end{center}
\end{figure}

The correlation between the difference of the spectral indices at
high and low frequencies ($\alpha_8^{20}-\alpha_5^8$) with
redshift (Fig.~\ref{fig:steepening_z}) shows a clear curvature in
the spectra. Note that the 20\,GHz flux densities from the higher
redshift objects correspond to flux densities from the higher
frequencies in the rest frame (scaling as $1+z$). Since the median
redshift of the QSO in the sample is 1.20, the steepening is
occurring at frequency $\nu> 50$\,GHz in the rest frame, and grows
steeper to above $\nu>70$\,GHz in the rest frame for the objects
at $z\sim 2.5$.

Although this correlation with redshift is most simply explained
as the combination of increased spectral curvature with frequency
and the changes in the rest frame frequency it should be noted
that the BSS sample does not cover a large enough flux range to
break the degeneracy between distance and power so it could also
be a correlation with power. Further investigation of this
correlation clearly needs the deeper sample, and would also
benefit from more complete redshift information, since there may
be selection effects in the sub-samples with existing redshift
information.

\section[]{Conclusions} \label{sec:Conclusions}
We have presented a complete sample of 320 sources selected within
the AT20G Survey catalogue as those having flux density $S_{20\rm
GHz}>0.50$\,Jy, $|b|>1.5^\circ$. Almost simultaneous 5 and 8\,GHz
observations have been used for spectral behaviour analysis.

Information on polarisation are available at all the frequencies.
We found that the median fractional polarisation is increasing
with frequency.

Neither the high frequency total intensity nor the polarisation
behaviour can be estimated from low frequency information. We
examined a set of issues that support this statement:
\begin{itemize}
\item the colour-colour plots show a broad range of spectral
shape: most sources spectra are not power-law so do not allow
easily extrapolation from one frequency to the other;
\item the comparison with low frequency selected samples showed
that by increasing the constraints on the low frequency sample the
number of low frequency objects recovered also at high frequency
increased, but that the completeness of the predicted high
frequency sample gets poorer.  It is necessary to fine tune the
conditions on the low frequency sample to obtain a good trade-off
between completeness and identification rate, but there is no way
to select a low frequency sample that guarantees that all the
sources will constitute a complete high frequency sample;
\item the polarisation spectral shape does not agree in all the
cases with the total intensity: the lack of knowledge on
polarisation properties, together with unpredictable polarisation
spectral behaviour make any forecast extremely difficult.
\end{itemize}

It is clear that actual high frequency
samples are better than trying to predict them from lower frequencies.

So, the Bright Source Sample constitutes an unprecedented
collection of information at 20~GHz, that will turn to be of
importance by itself and for any future observations at high radio
frequencies.

The whole AT20G Survey, in fact, will improve the radiosource
population knowledge to much lower flux densities.

This amount of information will be of crucial interest for the
next generation telescope, to provide good sample of calibrators,
and for the CMB targeted missions, as a test for point source
detection techniques, as a help in point-source removal in any
component separation exercise and as a list of candidate pointing,
flux and possibly polarisation calibrators.

Even with the relatively superficial analysis presented here, we
find interesting new physical effects from this sample:
\begin{itemize}
\item spectral steepening is common in this class of object;
\item the spectral steepening correlates with redshift, possibly due
to changing rest frame frequency; \item sources with spectral peaks in the GHz
range are common in this sample and have high depolarisation on
the low frequency side of the peak.
\end{itemize}

\section*{Acknowledgements}

MM and GDZ acknowledge financial support from ASI (contract Planck
LFI Activity of Phase E2) and MUR.

We gratefully thank the staff at the Australia Telescope Compact
Array site, Narrabri (NSW), for the valuable support. The
Australia Telescope Compact Array is part of the Australia
Telescope which is funded by the Commonwealth of Australia for
operation as a National Facility managed by CSIRO.

This research has made use of the NASA/IPAC Extragalactic Database
(NED) which is operated by the Jet Propulsion Laboratory,
California Institute of Technology, under contract with the
National Aeronautics and Space Administration.

This research has made use of data obtained from the SuperCOSMOS
Science Archive, prepared and hosted by the Wide Field Astronomy
Unit, Institute for Astronomy, University of Edinburgh, which is
funded by the UK Particle Physics and Astronomy Research Council.

We thank the referee for his useful comments and corrections.

\appendix

\section[]{Individual sources notes}\label{sec:IndSrc}

\begin{description}
\item[\emph{Table~2 source 61}]: PKS~0454$-$81 appears in
the scan maps, but the follow-up data were degraded by bad weather
and we didn't have the opportunity to re-observe it. For this
source we obtained a flux density measurement from its
observations as a secondary calibrator in October 2006.

\item[\emph{Table 2 source 92}] (PKS 0637-752) is a quasar with an
asymmetric jet seen in radio and Xray images (Schwartz et al.\
2000). The tabulated flux density is dominated by the core with
about 10\% in the 15 arcsec jet. It is one of the largest (100\
kpc) and most luminous jets known with properties similar to
3C273.

\item[\emph{Table~2 source 109}] (PMN~J0835$-$5953) has a highly
inverted radio spectra, with spectral index $\alpha_5^{20}=+0.88$,
but has no obvious optical counterpart. Although the Galactic
latitude is relatively low ($b=11^\circ$), the optical extinction
is only 1.1\,mag in the B band. The lack of optical ID suggests
this could be a distant radio galaxy rather than a QSO.

\item[\emph{Table~2 source 151}] (PKS~1143$-$696) is a
resolved double in the SUMSS image, and is also double in the
20\,GHz image. The SUMSS source is larger than the ATCA beam at
20\,GHz, suggesting that the measured flux density may be a lower
limit to the true value. The position of the low-frequency radio
centroid is slightly different from the AT20G position.

\item[\emph{Table~2 source 211}] (PKS~1548$-$79) is a
relatively nearby ($z=0.15$) galaxy with an unresolved radio
source which has a steep spectrum in our 5, 8 and 20\,GHz data.
The galaxy has strong optical emission lines, and has been studied
in detail by Tadhunter et al.\ (2001).

\item[\emph{Table~2 source 221}] appears to be one component of a
source (PKS~1622$-$29) which is double (component separation
$\sim1.5$~arcmin) in the NVSS image. Both components fall within
the ATCA 5~GHz beam, but the 20~GHz image is centred on the
eastern component and the other component falls outside the
primary beam. Our measured 20~GHz flux density is therefore an
underestimation of the total flux density.

\item[\emph{Table~2 source 258}] The AT20G source
(corresponding to PKS\,1932$-$46) is flagged as extended, and the
image appears to show a compact double.  The source is a
30\,arcsec double at 5\,GHz (Duncan \& Sproats 1992). The optical
position given in NED is associated with a $z=0.231$ galaxy at
(J2000) 19:35:56.5 $-$46:20:41, which is offset by 3.2\,arcsec
from the AT20G position but appears to be the correct ID.

\item[\emph{Table~2 source 273}] (PKS~2052$-$47) is a $z=1.5$
QSO which is also detected as both an X--ray and a gamma--ray
source. Since this source is an ATCA calibrator, its flux density
has been monitored at several epochs during 2002--7. The
calibrator data suggest that our AT20G observation of this object
in October 2004 took place during the declining stage of a flaring
phase, during which the flux density of the source changed
rapidly. This fast change in flux and polarisation properties is
clearly visible in our data, with the 20~GHz flux density
decreasing by a factor of 2.5 in two days.  This makes it
difficult to give a reliable value for the flux density and
fractional polarisation of this source.

\item[\emph{Table~2 source 292}] (PKS~2227$-$3952) is a
resolved triple in the SUMSS image. The low-frequency emission
extends somewhat beyond the 20~GHz ATCA beam, but the source is
not flagged as extended here, since the 20~GHz flux is dominated
by the core.

\item[\emph{Table~2 source 310}], flagged as extended, appears to
be the core of a well known and highly-extended radio galaxy
PKS\,2331$-$240. The optical ID is a galaxy at z=0.0477. The
extended flux is well outside of the primary beam used for these
observations and the flux densities listed correspond mainly to
the core.

\item[\emph{Table~2 source 319}], (PKS~2356$-$61)is a FRII galaxy
characterized by four bright regions of emissions that are
slightly asymmetric about the core (Burke et al.).
\end{description}

\bsp

\label{lastpage}

\end{document}